\documentclass[apl,twocolumn,showpacs,superscriptaddress,letterpaper]{revtex4-2}
\usepackage{graphicx,amsmath,amsfonts}
\usepackage{bm}
\usepackage{mathrsfs}
\usepackage{color}

\usepackage{dcolumn}
\usepackage[dvipsnames]{xcolor}
\usepackage{hyperref}
\hypersetup{
	colorlinks=true,
	linkcolor=blue,
	filecolor=magenta,
	urlcolor=blue,
	citecolor= blue,
}
\usepackage{MnSymbol}%
\begin{document}
	
	\title{Layer-dependent electronic structures and magnetic ground states of polar-polar $\rm{LaVO_3/KTaO_3}\, (001)$ heterostructures}
	
	\author{Shubham Patel}\email{spatelphy@iitkgp.ac.in}
	\affiliation{Department of Physics, Indian Institute of Technology, Kharagpur-721302, India}
	\author{Narayan Mohanta}
	\affiliation{Department of Physics, Indian Institute of Technology Roorkee, Roorkee 247667, India}
	\author{Snehasish Nandy}
	\affiliation{Theoretical Division, Los Alamos National Laboratory, Los Alamos, New Mexico 87545, USA}
	\author{Subhendra D. Mahanti}
	\affiliation{Department of Physics and Astronomy, Michigan State University, East Lansing, Michigan 48824, USA}
	\author{A Taraphder$^1$}\email{arghya@phy.iitkgp.ac.in}

	\date{\today}
	
	\begin{abstract}
		Employing a first-principles and model Hamiltonian approach, we work out the electronic properties of polar-polar LaVO$_3$/KTaO$_3$ (LVO/KTO, 001) heterostrctures, with up to six layers of KTO and five layers of LVO. Our analyses indicate the existence of multiple Lifshitz transitions (LTs) within the $t_{2g}$ bands, which can be fine-tuned by adjusting the number of LVO layers or applying gate voltage. Contrary to the experimental report, spin-orbit coupling is found to be negligible, originating solely from the Ta $5d_{xy}$-derived band of KTO, while the 5$d_{xz}$ and 5$d_{yz}$ bands are considerably away from the Fermi level while LVO overlayers having no role in it. Magnetic properties of the heterostructures, due to Vanadium ions, exhibit a pronounced sensitivity to the number of LVO and KTO layers. Our calculations indicate that the interlayer AFM, (so called A-AFM), is energetically most favorable. This is further supported by ground state energy calculations on extended $\sqrt{2}\times\sqrt{2}$ supercells. Moreover, we find that an insulator to metal transition at the interface requires four LVO layers, corroborating the experimental observation. The interfaces featuring ferromagnetic (FM) ground states turn out to be \textit{half-metallic} after the critical thickness is reached. Considerations of the magnetic interactions appear crucial for the experimentally observed critical thickness for metallicity.
		
	\end{abstract}
	
	\maketitle
	\section{Introduction}
	Interfaces of polar-nonpolar perovskite heterostructures are well-studied in literature owing to various applications using their physical properties such as the formation of a two-dimensional electron gas (2DEG), metal-insulator transition, magneto-transport, magnetism, and superconductivity \cite{ohtomo2004high,caviglia2010two,bert2011direct}. The next generation perovskite hetero-structure is polar-polar where due to ionic interactions, an intrinsic electric field is induced \cite{wadehra2020planar,wang2016creating}. The ability to tune the physical and chemical properties of a system by an external electric field along with the induced internal electric field is a major objective of electronic, optoelectronic and nano-electronic applications. 
	
	A wide variety of interfaces have been extensively investigated both experimentally and theoretically; in particular, a well-known polar-nonpolar $\rm{LaAlO_3/SrTiO_3}$ (LAO/STO) (001) interface has emerged as a promising candidate for last few decades due to its remarkable features. It exhibits LAO thickness-dependent transport properties, absent in its bulk counterparts \cite{thiel2006tunable,pentcheva2009avoiding,zhou2015interplay}. Other than the metal-insulator transition, the Rashba spin-splitting \cite{zhong2013theory,mohanta2015multiband}, the coexistence of ferromagnetism and superconductivity \cite{bert2011direct,li2011coexistence,mohanta2015multiband}, electronic reconstruction \cite{zhou2015interplay}, oxygen-vacancy effects \cite{kalabukhov2007effect}, topological superconductivity \cite{mohanta2014topological,scheurer2015topological}, and its influence on Lifshitz transitions \cite{nandy2016anomalous} are explored in these interfaces. Notably, a critical thickness of four unit cells of LAO is necessary to make the interface conducting \cite{pentcheva2009avoiding}. Compared to the polar-nonpolar interfaces, which form the 2DEG via polar discontinuity, the polar-polar oxide interfaces \cite{wang2016creating} are expected to produce higher charge carrier density and carrier mobility as there are, in principle, two electron-donor layers in this case transferring electrons to the interface to avoid polar discontinuity \cite{popovic2008origin}. 
	
	In a recent experimental report, such a polar-polar interface is formed to study its transport properties, in which the layers of LaVO$_3$ are stacked over KTaO$_3$ a substrate, forming a LaVO$_3$/KTaO$_3$ (LVO/KTO) interface in (001) direction \cite{wadehra2020planar}. A strong anisotropic transverse and longitudinal magnetoresistance (MR) was interpreted as the signature of strong spin-orbit coupling (SOC). In addition, the nature of observed anomalous MR (AMR) and planar-Hall effect (PHE) \cite{nandy2017chiral} was thought to come from Weyl fermions. The bulk KTaO$_3$ is a band insulator with an optical gap of 3.5 eV \cite{wemple1965some,jellison2006optical} crystallizing in a cubic symmetry with space group 221 (Pm$\bar{3}$n). It has a sizable SOC strength compared to SrTiO$_3$ coming from the Ta $5d$ orbitals \cite{nakamura2009electric,shanavas2014electric,shanavas2015overview}. $ABO_3$ oxides form alternative ($+/-$) charges in (AO) and BO$_2$ layers. In case of KTaO$_3$, the sequence is $\rm{(K^+O^{2-})^{-1}/(Ta^{5+}O_2^{2-})^{+1}}$, while it is $\rm{(La^{3+}O^{2-})^{+1}/(V^{3+}O_2^{2-})^{-1}}$ for LaVO$_3$. Unlike LAO/STO, in LVO/KTO the two materials form a $+/+$ interface of (TaO$_2$)$^{+1}$/(LaO)$^{+1}$ type, when terminated at TaO$_2$.
	
	Electron-electron correlations are usually considered important in $d-$electrons systems for the metal-insulator transition (MIT) and the experimentally observed band gap. However, this is not always the case. As discussed in recent theoretical studies \cite{varignon2019origin,malyi2023insulating}, microscopic degrees of freedom such as structural distortions and magnetic order actively participate in breaking local symmetries. Such symmetry breaking (SB) could indeed lead to MIT. Lowering the lattice symmetry can lower overall energy, and lead to a more stable state. In addition, formation of a heterostructure lowers the energy compared to the sum of energies of the participating bulk materials. This specific class of materials is in polymorphous crystal geometries, in which local and global symmetries are not the same, as discussed by Malyi \textit{et al.} \cite{malyi2023insulating}. As we will show, in LVO/KTO systems we have studied, even though the structural and magnetic SBs are important, we find that the stabilisation of the insulating phase requires electronic correlation via Hubbard-like $U$.
	
	To address the issue of a conducting interface, and consequently, the formation of 2DEG, we investigate the electronic properties of the polar-polar (001) LVO/KTO interface using first-principles density functional calculations. The cubic LaVO$_3$ (LVO), which is metallic with mostly V $3d$ orbitals at the Fermi level (FL), forms a conducting interface with KTO substrate even with a single layer of LVO. Although it is unclear from the experimental report which structure of LVO was used for the transport investigation, the charge distribution in the thin films of LVO and a later study suggest that a cubic LVO was likely used to form the LVO/KTO interface \cite{kakkar2022rashba}. We therefore perform our investigations with cubic LVO \cite{wold1954perowskite,jiang2006prediction} to match the stoichiometry and structure of KTO. In passing, we note that the orthorhombic phase of LaVO$_3$ is a Mott-Hubbard insulator \cite{bordet1993structural,de2007orbital}, undergoing a metal-insulator transition \cite{miyasaka2000critical,hotta2007polar}. The cubic perovskite LVO is a ferromagnetic metal with a total magnetic moment $\sim2\mu_{B}$, as reported by Rashid \textit{et al.} \cite{rashid2017theoretical}. To the best of our knowledge, no previous studies discuss the magnetic order in LVO-based heterostructures, except one experiment \cite{luders2009room}. There are also some theoretical studies dedicated to the cubic phase of bulk LVO, investigating the magnetism \cite{rashid2017theoretical} and electron-correlation using DMFT \cite{dang2013designing, dang2013theory}. Along with the non-magnetic calculations, we also report various possible magnetic orders with different number of layers of LVO in the LVO/KTO heterostructure.
	
	The remainder of the paper is organized as follows. In Sec. \ref{sec:compdetails} and \ref{sec:structure}, we describe the computational details and the interface geometry, respectively. In Sec. \ref{sec:results}, we present the non-magnetic electronic band structures with different layers of LVO and KTO, and then in Sec. \ref{sec:magnetism} we discuss the possible magnetic ordering in the LVO layers. Finally, in Sec. \ref{sec:conclusion}, we conclude with our findings.

	\section{\label{sec:compdetails}Methodology}
	We use density functional theory as implemented in Vienna \textit{ab-initio} Simulation Package ({\small{VASP}}) within the plane-wave pseudopotential method \cite{kresse1996efficiency,kresse1996efficient}. To approximate the electron-ion potential, we use the projected wave method \cite{blochl1994projector}, and the Perdew-Burke-Ernzerhof of the generalized gradient approximation \cite{perdew1996generalized} is used to approximate exchange-correlation effects, taking into account the on-site Coulomb interactions (GGA+$U$) for La, Ta and V within Dudarev's approach \cite{dudarev1998electron}. An empirical $U$ value equal to 7.5 eV is used for La \cite{pentcheva2008ionic}, while for Ta and V it is 5 eV \cite{cooper2012enhanced} and 3.4 eV \cite{assmann2013oxide}, respectively. The calculations are performed with a kinetic energy cut off of 530 eV, Brillouin zone sampling with a Gamma-centered 11$\times$11$\times$1 k-mesh, and energy convergence criteria of $10^{-6}$ eV. We fully relax the atomic positions until the force on each atom is less than 0.05 eV/\AA. To discard any spurious interaction, a 20 \AA\ vacuum is added between the periodic films. We further perform spin-polarized calculations to investigate the magnetic ground state of different combinations of LVO and KTO layers. We use the initial magnetic moment along the z-direction for V sites of different layers of LVO in parallel (for ferromagnetic order) and antiparallel (for antiferromagnetic order) directions. We calculate the spin-polarized density of states using the tetrahedron smearing method as implemented in {\small{VASP}}.
	
	\begin{figure}[!htb]
		\centering
		\includegraphics[scale=0.47]{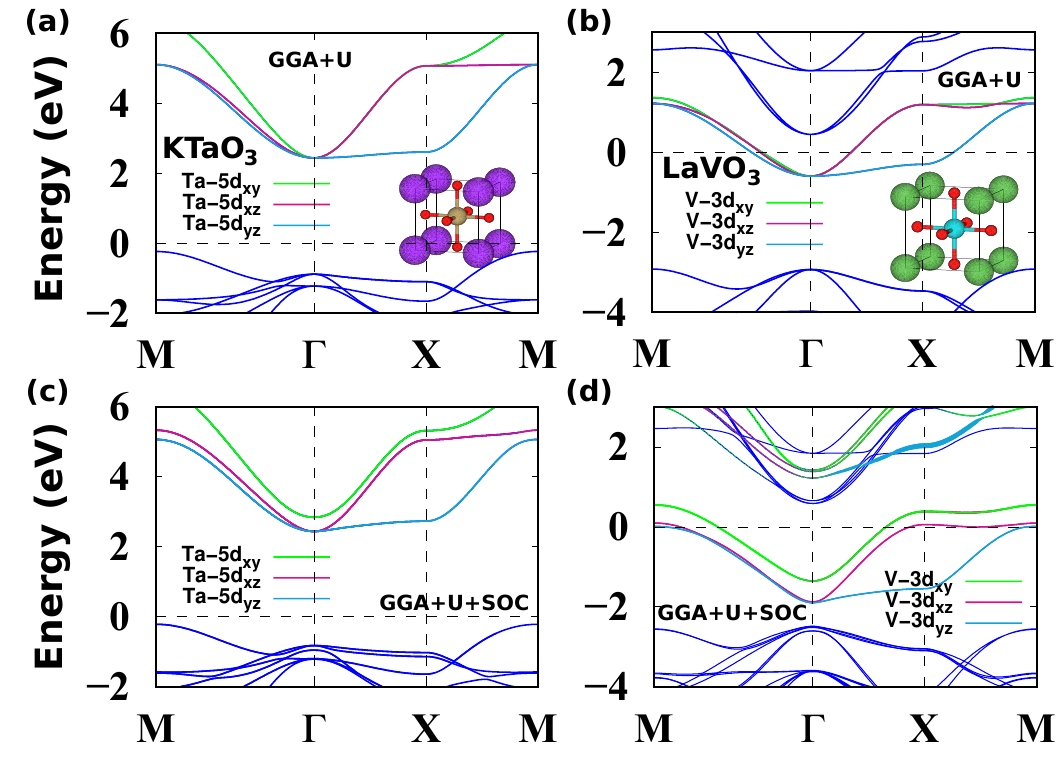}
		\caption{GGA+$U$ bulk band structures without SOC for the $ABO_3$-type cubic perovskites of (a) KTO and (b) LVO, with corresponding crystal structures shown in the insets. The large purple (green) spheres at the corners are the $A$ atoms, yellow (cyan) are the $B$ atoms at the center. The oxygen atoms are shown with small red spheres. SOC-incorporated bands are shown in the bottom panel in (c) for KTO and (d) for LVO. The green, magenta and cyan bands in all four figures present $d_{xy}$, $d_{xz}$ and $d_{yz}$ orbitals.}
		\label{fig:bulk}
	\end{figure}
	\begin{figure*}[!htb]
		\centering
		\includegraphics[scale=0.45]{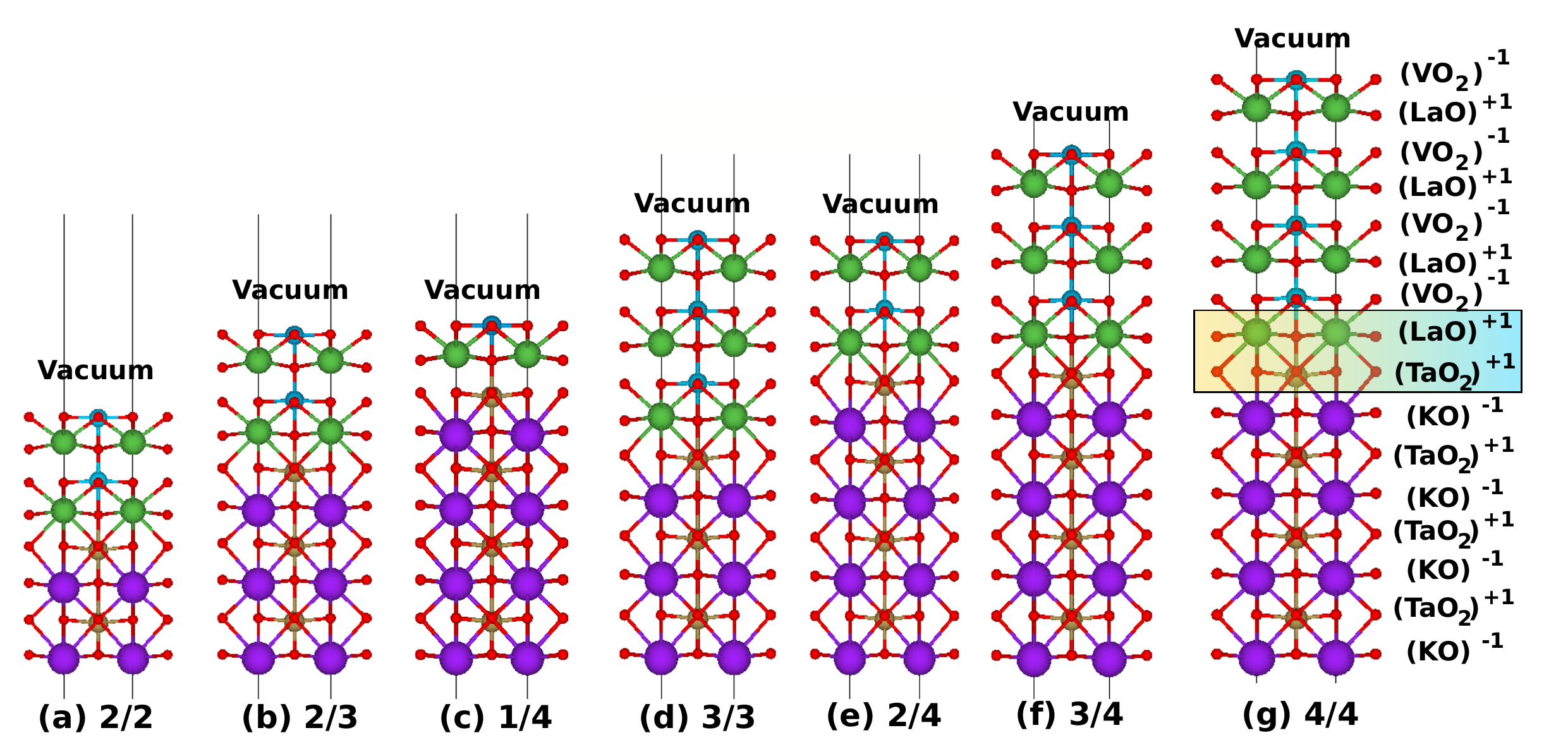}
		\caption{(a)-(g) (LVO)$_m$/(KTO)$_n$ heterostructure configurations with $m/n$ mentioned at the bottom of each. The interface is formed at the $\rm{(TaO_2)^{+1}}$ termination in each configuration, shown in (g)  with a shaded box. Also shown are the alternate charged layers.}
		\label{fig:interfaces}
	\end{figure*}

	\section{\label{sec:structure}Crystal structures and interfaces}
	From our calculations we find that bulk KTO is an indirect band gap insulator with a  gap $\sim2.7$ eV. It is found to be in the cubic phase with space group 221 (Pm$\bar{3}$n) (see Fig. \ref{fig:bulk}(a)) and lattice parameters $a=b=c = 4.03 $ \AA\, which are in good agreement with the previously reported values \cite{wemple1965some,jellison2006optical}. The corresponding band structure is presented in Fig. \ref{fig:bulk}(a). The cubic phase of bulk LaVO$_3$ (LVO), on the other hand, shows metallic behavior. The calculated lattice parameter ($a$) of LVO is 3.95 \AA\, in good agreement with the experimentally reported value (3.91 \AA\, \cite{wold1954perowskite,jiang2006prediction}). Naturally occurring LVO has an orthorhombic crystal structure and is an insulator with a 1.1 eV band gap \cite{bordet1993structural,de2007orbital,miyasaka2000critical}.  Including SOC, we note that the degeneracy between the $t_{2g}$ bands ($d_{xy}$, $d_{xz}$ and $d_{yz}$) is lifted in both KTO and LVO as shown in Figs. \ref{fig:bulk}(c) and (d), respectively. We use a cubic phase \cite{wold1954perowskite} of LVO for the hetero-structure with cubic KTO as a substrate to match the structures, which was most likely the case for experimental haterostructures.
	
	To prepare the heterostructures (LVO)$_m$/(KTO)$_n$, we use different layers ($m$ and $n$) of LVO and KTO. We terminate the KTO unit cell at $\rm{(TaO_2)^{+1}}$ layer and LVO at $\rm{(LaO)^{+1}}$ layer to form a polar-polar interface, $\rm{(LaO)^{+1}}$/$\rm{(TaO_2)^{+1}}$. To compare the results and see the effect of the number of layers of (LVO)$_m$ and (KTO)$_n$, we use various configurations as illustrated in Fig \ref{fig:interfaces}. It is important to note that the crystal symmetry of the supercell structures are reduced from the bulk cubic structure to tetragonal P4mm with space group No. 99. All the defined geometries are relaxed with the conjugate gradient algorithm as implemented in the {\small{VASP}} package. While optimizing the geometries, we kept the in-plane lattice constant equal to 4.03\AA\, for all the supercells, which is also the lattice constant for bulk KTO. We are using KTO as substrate - the LVO layers are under a tensile strain of 1.98\%. We check the lattice constant with full geometry relaxation (cell parameters and atomic positions) for the smallest supercell, 2/2. The lattice constant with fully relaxed structure for 2/2 is 3.99\AA\,, that is an average of the lattice constants of LVO and KTO. In Sec. \ref{sec:magnetism}, we will discuss in detail the difference in the electronic structures of this fully relaxed 2/2 and laterally constrained 2/2. Here we just point out that the two sets of results are in qualitative agreement.

   In order to discuss the stability and the possibility of the formation of the heterostructures in any given crystal structure, we calculate the formation energies for each of the configurations. The formation energy is defined as:  $E_{f[(LVO)_m/(KTO)_n]} = E_{(LVO)_m/(KTO)_n}- E_{(LVO)_m}-E_{(KTO)_n}$, where $E_i$ are the ground state energies of the $i$th system. It is energetically favorable to form the heterostructures, indicated by the negative formation energies outlined in Table T1 of SM \cite{patel2023layer}. 
	
	\begin{figure*}[!htb]
		\centering
		\includegraphics[scale=0.65]{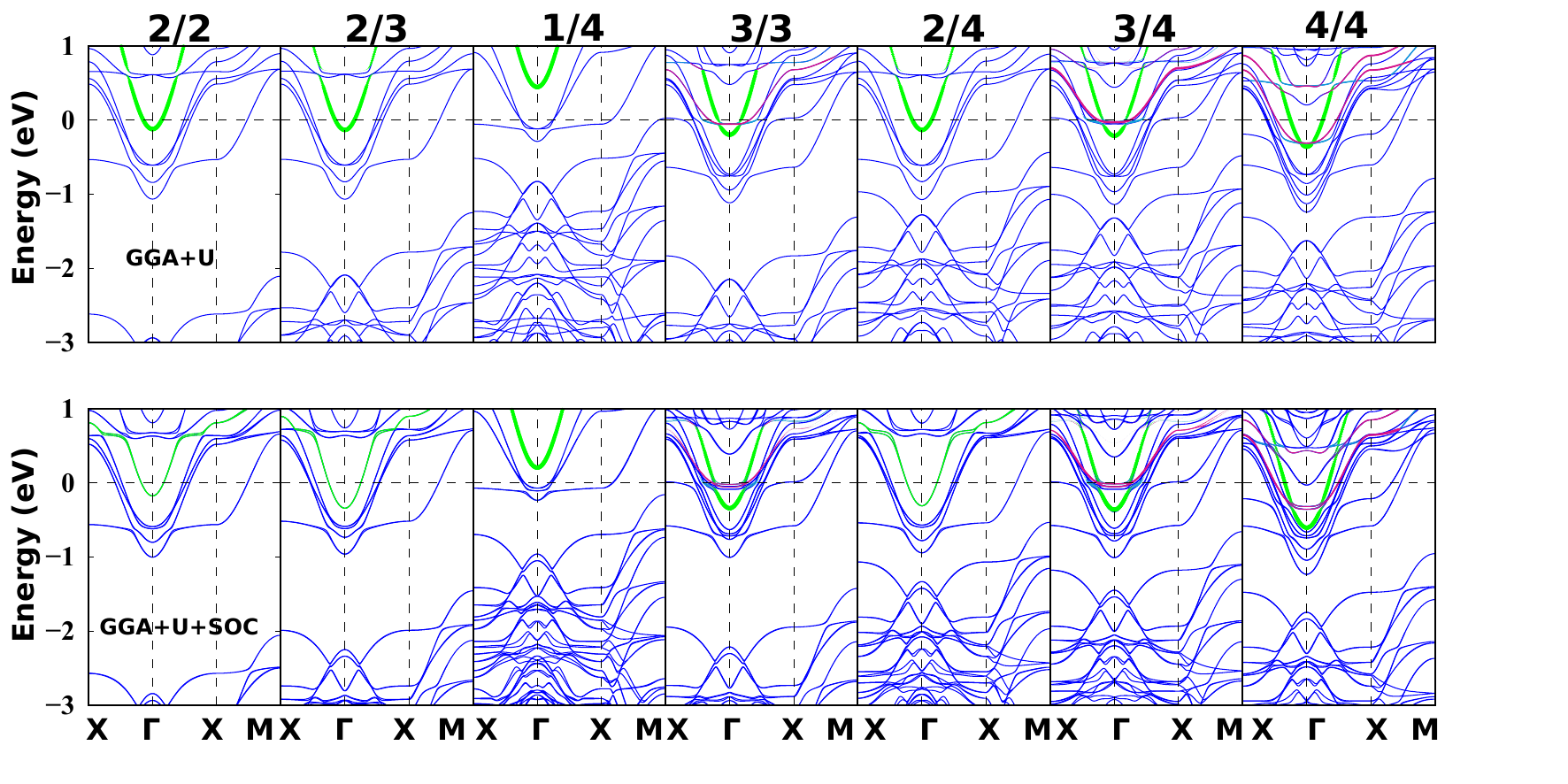}
		\caption{Non-magnetic band structures without (upper panel) and with (bottom panel) SOC for various configurations shown in Fig. \ref{fig:interfaces} along X---$\Gamma$---X---M high symmetry path of a square Brillouin zone. The bands in green, magenta and cyan are $\rm{Ta_{int}}$ $5d_{xy}$, V $3d_{xz}$ and V $3d_{yz}$, respectively. Lifshitz transition is shown for 3/3, 3/4 and 4/4 configurations with green ($\rm{Ta_{int}}$), magenta (V $3d_{xz}$) and cyan (V $3d_{yz}$) bands. Both atomic and Rashba SOC are weak for all the cases. SOC breaks the degeneracy between the $3d_{xz}$ and V $3d_{yz}$ orbitals weakly in the bands involved in Lifshitz transition.}
		\label{fig:bands}
	\end{figure*}

	\section{\label{sec:results}Non-magnetic electronic structure}
	
	In the following, we perform layer-dependent electronic structure calculations for different heterostructures given in Fig. \ref{fig:interfaces}. We show the band structures with non-magnetic calculations in Fig. \ref{fig:bands}. The cubic LVO being metallic, a single layer of LVO on the KTO substrate is enough to make the interface conducting. From the non-magnetic calculations, we observe    

	\begin{itemize}
		\item[$\blacksquare$] Increasing KTO layers only shifts the oxygen bands in the valence band region towards the FL but does not affect the conduction bands. The band-widths of the systems with the same number of layers of LVO ($m$) are almost the same for any $n$, indicating the band-width is $n$-independent, where $n$ is the number of KTO layers. 
		
		\item[$\blacksquare$] Though all the surfaces under investigation are conducting, including single LVO layer, as we will discuss shortly in Sec. \ref{sec:proj_bands} of projected bands, more than one layer of LVO (i.e., $m > 1$) is needed for Ta$_{\rm{int}}\,5d_{xy}$ orbital to take part in charge transport, where $\rm{Ta_{int}}$ is the Ta atom at the interface.
		
		\item[$\blacksquare$] The interfaces with three and more than three layers of LVO (i.e., $m \ge 3$) show a Lifshitz transition  \cite{zhong2013theory,nandy2016anomalous} near the FL as shown in Fig. \ref{fig:bands}. It is important to mention that the Lifshitz transition is not coming from the $t_{2g}$ manifold of the interfacial transition metal Ta. Unlike LAO/STO (Ti $t_{2g}$ at the interface), the Lifshitz transition in the present case is formed between the interfacial Ta $d_{xy}$ band and the V $d_{xz,yz}$ (near the interface) bands. We will have a clear idea from the projected band structure in Fig. \ref{fig:proj_bands} of Sec. \ref{sec:proj_bands}. It is well understood that the Lifshitz transition plays a vital role in anomalous transport, such as thermal conductivity, as well as the Seebeck coefficient. The change in the Fermi surface topology, induced by Lifshitz transition, gives rise to a cusp in the dc and thermal conductivity as reported by Nandy \textit{et al.} \cite{nandy2016anomalous}. Our results involving multiple layers of LVO/KTO indicate that the Lifshitz transition point can be adjusted by varying the number of LVO layers, eliminating the need for external gate voltage. This feature makes these heterostructures particularly suitable for application in thermal transport devices. Among these structures, those with three layers of LVO emerge as optimal candidates due to the proximity of the Lifshitz point to the FL.

		\item[$\blacksquare$]  Contrary to experimental claim \cite{wadehra2020planar}, there is hardly any SOC in the bands involved in transport at the interface: the bands near the FL show negligible SOC. The small SOC breaks the degeneracy slightly at the Lifshitz points. The Rashba spin-splitting is negligible in all the cases. Contrary to the large SOC (in the Ta $5d$ orbitals) in bulk KTO, as shown in Fig. \ref{fig:bulk}(c), the LVO/KTO interfaces studied here do not show significant SOC. A recent report, however, claims significant SOC in SrTaO$_3$/KTaO$_3$ interface \cite{zhang2023quasi} having Ta on both the materials. 
		
	\end{itemize}

	A bare KTO slab was then checked for a strong Rashba spin-splitting. We calculate the band structure of a thin film constructed with eight unit-cells of KTO (shown in Fig. S1(a) of SM \cite{patel2023layer}) and find that the strength of Rashba SOC is fairly large. In Fig. S1(b) of SM \cite{patel2023layer}, we see that there are several bands around the $\Gamma$ point, with considerable Rashba spin-splitting (shown inside a red rectangle). The bands with strong Rashba spin-splittings are around 0.5 eV above the Fermi energy, while the Rashba spin-split bands for the 4/4 interface are far removed from the FL, around 1.2 eV above it (see Fig. \ref{fig:bands} and Fig. \ref{fig:proj_bands}(e)). We wish to emphasize that the spin-split bands observed in Fig. \ref{fig:proj_bands}(e) arise from the Ta $5d_{yz}$ and Ta $5d_{zx}$ orbitals, and not from the Ta $5d_{xy}$ orbital. It is probable that increasing the number of KTO layers (beyond 8) could bring these spin-split bands closer to the FL, as reported in Kakkar \textit{et al.} \cite{kakkar2022rashba}. The conduction band topology is unaffected with increasing number of KTO layers for the heterointerface LVO/KTO (see Fig. \ref{fig:bands}) though. It seems that the substantial Rashba spin-orbit coupling primarily originates from the KTO surface rather than the LVO layers in the LVO/KTO interfaces. Few layers of KTO have minimal Rashba SOC to show up in transport.

	\subsection{\label{sec:layerdos} Layer projected density of states}
	To assess the role of individual layers within the slab, we examine the layer-projected DOS for the 4/4 LVO/KTO hetero-interface, as illustrated in Fig. S1(c) of SM \cite{patel2023layer}. The layer-projected DOS analysis reveals that, among all the Ta atoms in the KTO substrate, only the interfacial Ta atom (Ta$\rm_{int}$) participates in charge transport. Conversely, all the V atoms in the slab contribute to the FL. Our layer-projected DOS highlights a nearly equal contribution from all V atoms at the FL.
	
	\subsection{\label{sec:proj_bands}Projected band structure}
	
\begin{figure*}[!htb]
		\centering
		\includegraphics[scale=0.7]{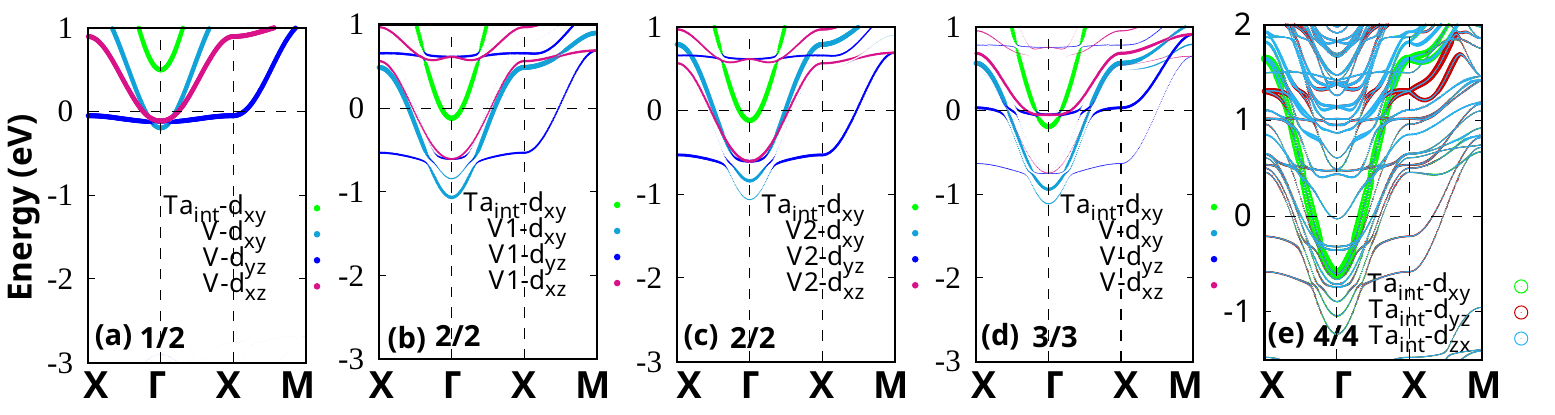}
		\caption{The orbital-projected bands (without SOC) of Ta$_{\rm{int}}\, 5d_{xy}$ at the interface (green), and $t_{2g}$ of V (blue and magenta) just above the interface. V1 and V2 in (b) and (c) represent the nearest and farthest V atoms from the interface, respectively. (d) The bands shown in green, blue and magenta circles participate in the Lifshitz transition near the FL in a 3/3 interface. (e) The projected bands with SOC for 4/4 case. The Rashba spin-split bands, contributed by Ta $d_{yz,zx}$, are about 1.2 eV above the FL. }
		\label{fig:proj_bands}
	\end{figure*}

	To gain insight into the distinct orbital contributions at the interface, we illustrate orbital-projected bands without SOC in Fig. \ref{fig:proj_bands}. Fig. \ref{fig:proj_bands}(a) reveals that, in the context of 1/n (one layer of LVO and n layers of KTO), only the $t_{2g}$ band manifold of V $3d$ bands is contributing near the FL, with the interfacial Ta playing no role in transport. The green band is the $d_{xy}$ orbital of the interfacial Ta atom. Our analysis indicates that more than one layer of cubic LVO is necessary to bring the interfacial Ta 5d$_{xy}$ band down to the FL. For the 2/2 configuration (refer to Figs. \ref{fig:proj_bands}(b,c)), both the $d_{xz,yz}$ bands from the nearest and farthest V atoms contribute nearly equally.  As previously noted, a Lifshitz transition is evident in cases involving three or more layers of LVO, as shown in Fig. \ref{fig:proj_bands}(d), where a Lifshitz transition occurs between Ta$_{\rm{int}},d_{xy}$ and V $d_{xz,yz}$ bands. Here, the interfaces with three layers of LVO are intriguing due to their Lifshitz point being closest to the FL, as observed in \ref{fig:proj_bands}(d). This characteristic, presumably, makes the 3/n heterostructures ideal candidates for thermoelectric applications. The Rashba spin-split bands for the $4/4$ system are depicted in Fig. \ref{fig:proj_bands}(e) with red and blue circles. It is important to reiterate that the Rashba spin-splitting in the Ta $d_{xy}$ band is minimal, while a significant contribution to Rashba SOC strength is observed in the Ta $d_{yz,zx}$ bands approximately 1.2 eV above the FL.
	
	\subsection{\label{sec:model}Model Hamiltonian}
	
	To conduct a theoretical analysis of physics near the Lifshitz point, we formulate an effective Hamiltonian model. In the absence of SOC, we specifically select three bands in close proximity to the Fermi energy (refer to Fig. \ref{fig:proj_bands}(d)). These bands originate from the contributions of Ta $d_{xy}$ and V $d_{xz,yz}$. The corresponding Hamiltonian model for these three bands is expressed through Slater-Koster integrals, defined as follows.
	\begin{center}			
		$$t_{xy,xy} = 3m^2n^2\,t_{dd\sigma} + (m^2+n^2-4m^2n^2)\,t_{dd\pi} + (l^2+m^2n^2)\,t_{dd\delta}$$
		\begin{equation}	
			t_{yz,yz} = 3l^2m^2t_{dd\sigma} + (l^2+m^2-4l^2m^2)\,t_{dd\pi} + (n^2+l^2m^2)\,t_{dd\delta}
		\end{equation}	
		$$t_{zx,zx} = 3n^2l^2t_{dd\sigma} + (n^2+l^2-4n^2l^2)\,t_{dd\pi} + (m^2+n^2l^2)\,t_{dd\delta}$$
	\end{center}
	where, $l$, $m$ and $n$ are the direction cosines of Ta atom at the interface and V atom near the interface, and $t_{dd\sigma}$, $t_{dd\pi}$, and $t_{dd\delta}$ are the Slater-Koster overlap parameters. 
	
	We construct the three-band model Hamiltonian, 
	\begin{center}
		\begin{equation}
			\mathcal{H}_0 = 
			\begin{pmatrix}
				h_1 & h_{12} & h_{13}\\
				h_{21} & h_2 & h_{23}\\
				h_{31} & h_{32} & h_{3}
			\end{pmatrix} 
			= 
			\begin{pmatrix}
				h_1 & 0 & 0\\
				0 & h_2 & 0\\
				0 & 0 & h_{3}
			\end{pmatrix} 
		\end{equation}
	\end{center}
	with $h_1$, $h_2$ and $h_3$ being $d_{xy}$, $d_{yz}$ and $d_{zx}$ band dispersion relations, respectively. The off-diagonal elements are zero due to symmetry of the orthogonal orbitals. These diagonal matrix elements are constructed using the Slater-Koster integrals as we present in Eqn. \ref{eq:SK}.
	
	\begin{center}
		$$ h_1 =\epsilon_1 - 2t_{1}(cos(k_x) + cos(k_y)) -t_2 -4\,t_3\, cos(k_x)\,cos(k_y) $$
		\begin{equation}\label{eq:SK}
			h_2 = \epsilon_2 - 2(t_{1}\, cos(k_x) + t_{2}\,cos(k_y)) - t_1 - 2\,t_3 \,cos(k_y)
		\end{equation}
		$$ h_3 = \epsilon_3 - 2(t_{2}\, cos(k_x) + t_{4}\,cos(k_y)) - t_4 - 2\,t_5 \,cos(k_x)$$ 
	\end{center}
	where, $\epsilon_i$ are the onsite energies for Ta $d_{xy}$ and V $d_{xz,yz}$ orbitals, and $t_{i}$'s are the hopping integrals. The $t_{2g}$ bands around $\Gamma$ point are present in Fig. S2 of SM \cite{patel2023layer}. The strong hybridization between three orbitals leads to avoided band crossings around the $\Gamma$ point. This avoided crossing also represents a broken symmetry which is the inversion symmetry in the case of the interface. We incorporate the broken inversion symmetry in the tight-binding Hamiltonian using atomic SOC and the orbital Rashba effect. The atomic SOC and the orbital Rashba Hamiltonian, in basis of [Ta $d_{xy,\sigma}$, V $d_{yz,\sigma}$,V $d_{zx,\sigma}$], $\sigma$ is the spin index, can be defined in the following manner \cite{mohanta2015multiband},
	
	\begin{center}
		\begin{equation}
			\mathcal{H}_{ASO} = 
			\frac{\zeta}{2}\begin{pmatrix}
				0 & 0 & 0 & 0  & 1 & -i\\
				0 & 0 & i & -1 & 0 &  0\\
				0 & -i & 0 & i  & 0 &  0\\
				0 & -1 & -i & 0 & 0 & 0 \\
				1 & 0 & 0 & 0 & 0 & -i\\
				i & 0 & 0 & 0 & i & 0\\
			\end{pmatrix} ,
		\end{equation}
		\begin{equation}
			\mathcal{H}_{RSO} = 
			\gamma\begin{pmatrix}
				0 & -2i\,sin(k_x) & -2i\,sin(k_y) \\
				2i\,sin(k_x) & 0 & 0\\
				2i\,sin(k_y) & 0 & 0
			\end{pmatrix} 
		\end{equation}
		
	\end{center}
	where, $\zeta$ and $\gamma$ are the atomic and Rashba SOC, respectively. The effective Hamiltonian including spin-orbit interactions is $\mathcal{H} = \mathcal{H}_0 + \mathcal{H}_{ASO} + \mathcal{H}_{RSO}$. Fitting this six-band Hamiltonian near the $\Gamma$ point gives us the unknown parameters, which are as follows (in eV): $\epsilon_1 = 0.76$, $\epsilon_2 = 0.7$, $\epsilon_3 = 0.57$, $t_1 = 0.0919$. $t_2 = 0.1778$, $t_3 = 0.2048$, $t_4 = 0.1186$, $t_5 = 0.1295$, $\zeta = 0.03$, and $\gamma = 0.015$. The fitted bands using these parameters are presented in Fig. S2 of SM \cite{patel2023layer}.

\begin{figure*}[!htb]
		\centering
		\includegraphics[scale=0.55]{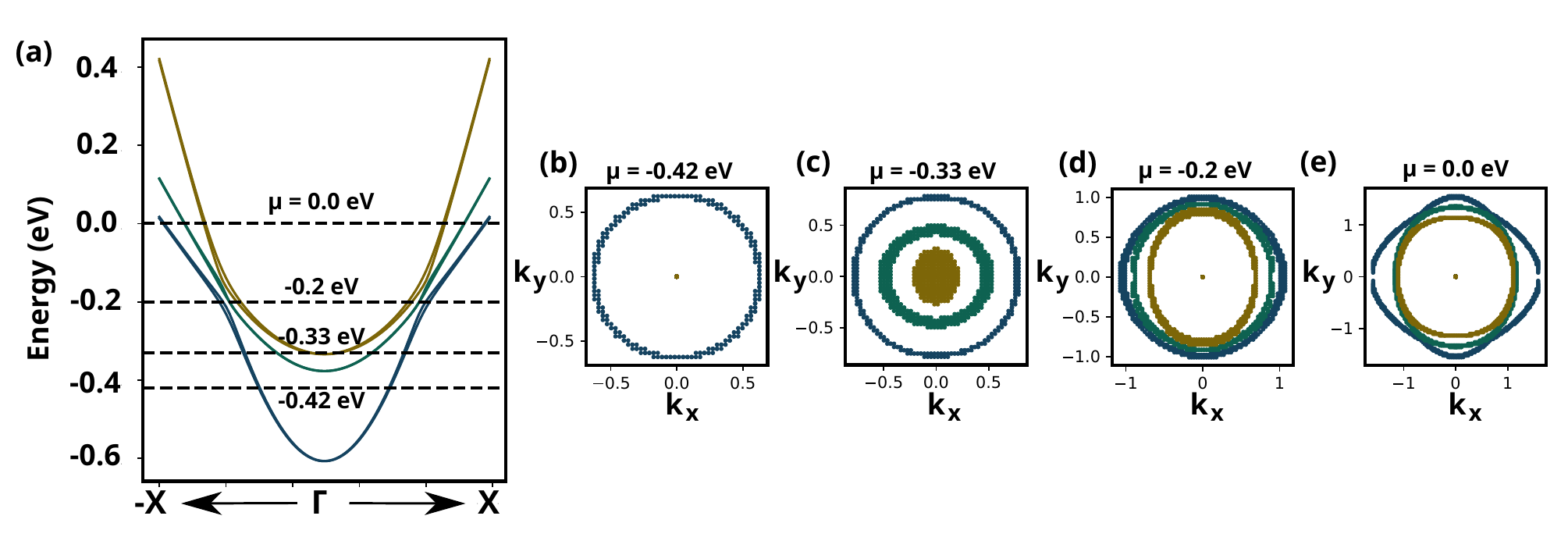}
		\caption{(a) The tight-binding band structure and (b)-(e) the change in the Fermi surface topologies can be observed at different chemical potentials ($\mu$). The Fermi surfaces are plotted in the $k_x - k_y$ plane and in units of $\pi$. }
		\label{fig:fsurf}
	\end{figure*}

	In Fig. \ref{fig:fsurf}, we depict the Fermi surface, illustrating its transformation under varying chemical potentials ($\mu$). The Lifshitz transition, denoting a shift in the Fermi surface topology, can be controlled experimentally by adjusting the chemical potential, carrier density, and, consequently, the Lifshitz point through gate voltage modulation. Notably, at $\mu = -0.42$ eV, only two orbitals form electron-like pockets, but as $\mu$ increases, additional electron-like pockets emerge as shown in Fig. \ref{fig:fsurf}. At $\mu = -0.33$ eV, all three pockets contribute to the Fermi surface. This highlights the $3/3$ configured hetero-interface as an optimal choice for transport devices due to its proximity to the FL.
	
	\section{\label{sec:magnetism}Magnetic ordering}
	
\begin{figure}[!htb]
		\centering
		\includegraphics[scale=0.6]{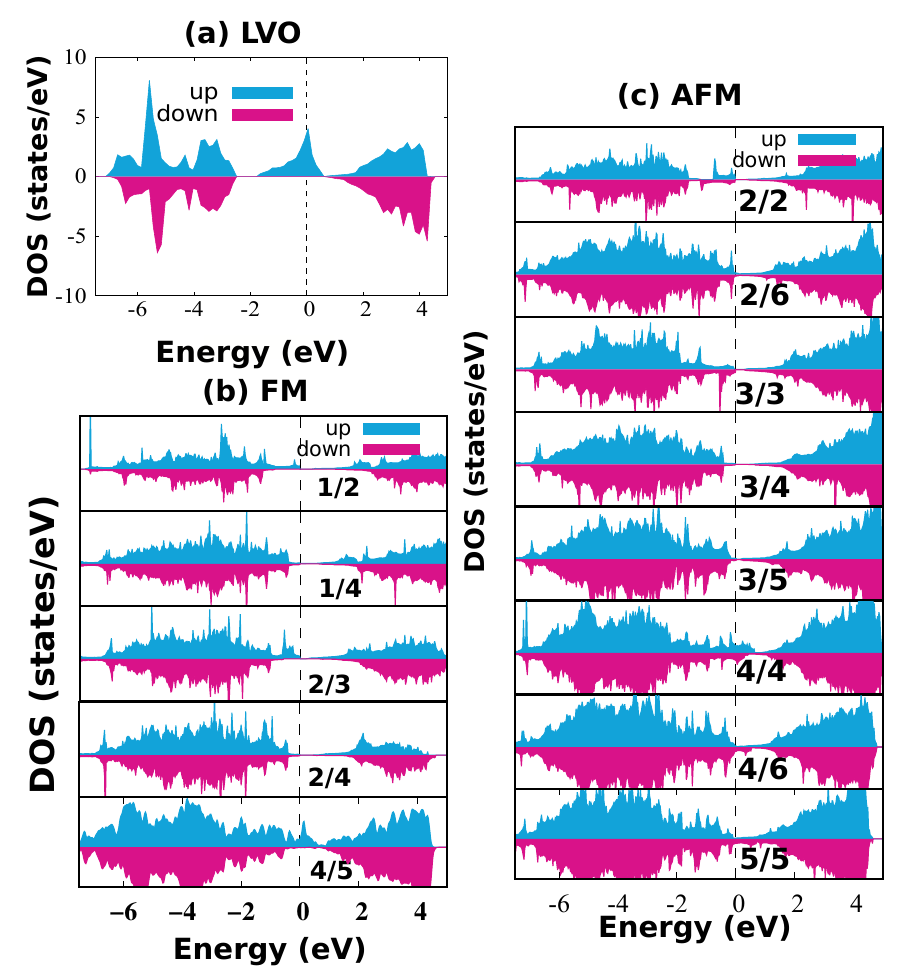}
		\caption{(a) The spin-resolved density of states (DOS) for the cubic LVO. The gap in the minority spin channel shows it is an FM \textit{half-metal}. (b) The FM DOS for various interfaces. (c) DOS for the AFM order. The DOS range for all the plots is $-20$ to $25$ states/eV. }
		\label{fig:sp-dos}
	\end{figure}

	Numerous investigations are available in literature for the magnetic characteristics of bulk LVO compound \cite{bordet1993structural, sawada1996orbital,fang2003anisotropic,fang2004quantum, mahajan1992magnetic}. With a total spin moment $S=1$ from V$^{3+}$ ionic configuration (3d$^2$), vanadium (V) plays a crucial role in the magnetism of LVO. The pseudo-cubic phase of LVO, with a tetragonal distortion of the cubic structure, exhibits C-type antiferromagnetic (AFM) order at low temperatures. In the C-type AFM, V moments in the basal plane (ab plane) are antiferromagnetically ordered while the moments along the c-axis are ordered ferromagnetically. The susceptibility displays a peak at the transition temperature $T_N \sim$ 140K and there is a concomitant orthorhombic to monoclinic transition at that temperature \cite{bordet1993structural, mahajan1992magnetic}. Conversely, the cubic LVO, sharing the same space group as KTO (Pm$\bar{3}$n), is claimed to be ferromagnetic (FM) \cite{rashid2017theoretical}. In Fig. \ref{fig:sp-dos}(a), we present the spin-polarized DOS for bulk cubic LVO, revealing the unmistakable FM nature of the cubic phase. Notably, the DOS illustrates a substantial band gap (approximately 4 eV) in the spin-down channel, while the spin up channel shows finite DOS at FL, indicative of a \textit{half-metal}. This was reported earlier as well \cite{rashid2017theoretical}. The \textit{half-metallic} character, a distinctive feature of the ferromagnetic state \cite{de1983new}, has its potential utility in spintronics devices, where both spin current and charge current are harnessed \cite{picozzi2003first,mavropoulos2005half,hashemifar2005preserving,di2018half,jena2023evidence,spinelectronics}.
	
	To explore the magnetic order in the (001) interface composed of LVO/KTO, we conducted spin-polarized density functional theory calculations for a range of interfaces illustrated in Fig. \ref{fig:interfaces}. The sublattice magnetization was constrained to exhibit parallel alignment on V atoms across different layers for FM order and antiparallel alignment for AFM order. The crystal structures are relaxed with different magnetic orderings, by keeping again the in-plane lattice constant fixed. The relaxed geometries are shown in Fig. S3(b) of SM \cite{patel2023layer}. After relaxation there is a uniform elongation of the atomic layers in the magnetic case, making the system more stable compared to the non-magnetic case (Fig. S3(a) of SM \cite{patel2023layer}), where the LVO and KTO layers are non-uniformly deforming the overall geometry. This uniform relaxation of atomic positions in the magnetic case is possibly the reason for the insulating phases, which we are going to discuss shortly. Moreover, we also perform a full geometry relaxation (atomic positions and lattice parameters), and find that the electronic structure of the system is unaltered, implying the local relaxations are more important than the global geometry relaxation. The AFM DOS for 2/2 with partial (atomic positions) and full geometry relaxation are shown in Fig. S4 of SM \cite{patel2023layer}. We then compared the spin-polarized ground state energies (GSEs) of all the heterostructures, considering various magnetic orderings such as non-magnetic (NM) with zero moment, FM, and AFM. By AFM we mean that each V layer is ferromagnetic but the layers are antiferromagnetically ordered as one goes from one layer to the next in the (001) direction. In a bulk system this type of AFM ordering is referred to as A-AFM. The resulting data are organized in Table \ref{table:magnetic}. Notably, we observe that energetically the AFM ground state consistently emerges as the most favorable ground state in all the heterostructures excepting for the monolayer of LVO. An examination of Table \ref{table:magnetic} reveals a small difference (approximately 0.1 eV) between GSEs of FM and AFM with equal number of LVO and KTO layers, suggestive of a potential coexistence of AFM and FM ground states \cite{zhang2020coexistence}, and a possible transition between them. To assess the stability of the magnetic behavior with variations of Hubbard-$U$, we perform the spin-polarized calculations for some of the interfaces with a higher $U$ of 5 eV. Assuredly, the ground state remains nearly insensitive to changes in $U$, as evidenced by the GSEs detailed in Table \ref{table:magnetic}.

	\begin{center}
		\begin{table*}[!ht]
			\centering
			\setlength{\tabcolsep}{20pt} 
			\renewcommand{\arraystretch}{1.2}
			\caption{\label{table:magnetic} The ground state energies (in eV) for non-magnetic (NM), ferromagnetic (FM) and Antiferromagnetic (AFM) ordering. FM implies both intraplanar (parallel to the interface) and interplanar (perpendicular to the interface) spin orientations, and AFM implies intraplanar ordering is ferromagnetic and interplanar nearest neighbor ordering is antiferromagnetic (commonly known as A-AFM in 3D magnetically ordered systems). The value of $U$ used for V 3d orbitals is 3 eV (the cases with $U$ = 5 eV are shown in the bracket).}
			\begin{tabular}{ c c c c c}
				\hline
				\hline
				interface &	NM (eV) &	FM (eV) & AFM (eV) & Nature\\
				\hline
				1/2 	& -112.212 & -113.655 & & FM-Insulator\\
				1/4 	& -186.148 & -187.592 & & FM-Insulator\\
				2/2 	& -151.557 & -153.938 (-151.952) & -153.989 (-151.985) & AFM-Insulator\\
				2/3  	& -188.520 & -190.907 & -190.959 & AFM-Insulator\\
				2/4 	& -225.475 & -227.868 & -227.923 & AFM-Insulator\\
				2/6 	& -299.447 & -301.853 & -301.896 & AFM-Insulator\\
				3/3 	& -227.694 & -230.728 & -231.187 & AFM-Insulator\\
				3/4  	& -264.643 & -266.855 (-264.568) & -267.794 (-265.085) & AFM-Insulator\\
				3/5 	& -301.649 & -304.962 & -305.166 & AFM-Insulator\\
				4/4 	& -303.897 & -307.732 & -308.364 & AFM-Metal\\
				4/5 	& -340.926 & -345.012 & -345.082 & AFM-Metal\\
				4/6	& -375.535 & -380.743 & -382.245 & AFM-Metal\\
				5/5	& -380.247 & -385.253 & -385.732 & AFM-Metal\\
				\hline
				\hline
			\end{tabular}
		\end{table*}
	\end{center}
	
	As we noted earlier in the introduction, the local symmetries of bulk materials could be lowered by forming heterostructures resulting in MIT \cite{varignon2019origin,malyi2023insulating}. If the energy can be lowered by lowering symmetries, it indicates a possible stable phase and an electronic phase transition might occur. By forming interfaces and introducing magnetic order, we break crystal symmetries. We calculated the relative GSEs per formula unit for each configuration and found that the energies are lowered for magnetic states. The measured relative GSEs are tabulated in Table T2 of SM \cite{patel2023layer}. We find that the AFM and FM are more stable than the non-magnetic phases. However, even though we have symmetry breaking in the system, it is not enough to open a gap at the FL in the single particle spectrum. In order to open a band gap, addition of Hubbard-$U$ in the density functional calculations is required (in addition to local structural and magnetic SBs). For example, we check the AFM-DOS without and with $U$ for the 2/2 LVO/KTO heterostructure (see Fig. S5 of SM \cite{patel2023layer}), and find that effect of electron-electron correlation is necessary for a gap to open up between occupied and unoccupied single particle states.

	Examining the ferromagnetic DOS (though it is less stable compared to AFM) shown in Fig. \ref{fig:sp-dos}(b), it is evident that the interfaces exhibit a \textit{half-metallic} behavior. Specifically, only one spin channel conducts at the FL, while the other channel remains gapped. All FM LVO/KTO (001) interfaces appear to possess inherent \textit{half-metallic} character beyond three LVO layers. It is crucial to highlight that the bulk LVO itself is \textit{half-metallic}, but this \textit{half-metallicity} is suppressed in (LVO)$_m$/(KTO)$_n$ heterostructures, where, $m<4$. In these cases, the system makes a transition to an insulating state. It requires more than three layers of LVO for the bands of one of the spin channels to appear at the FL in the FM heterostructure, and make it conducting (as \textit{half-metal}). This is illustrated in Fig. \ref{fig:sp-dos}(b). Turning attention to the AFM cases in Fig. \ref{fig:sp-dos}(c), it is seen that they exhibit insulating behavior for configurations with fewer than four layers of LVO. Additionally, the magnetic moment on each V atom is found to be slightly less than 2$\mu_{B}$.
	
	\begin{figure*}[!htb]
		\centering
		\includegraphics[scale=0.36]{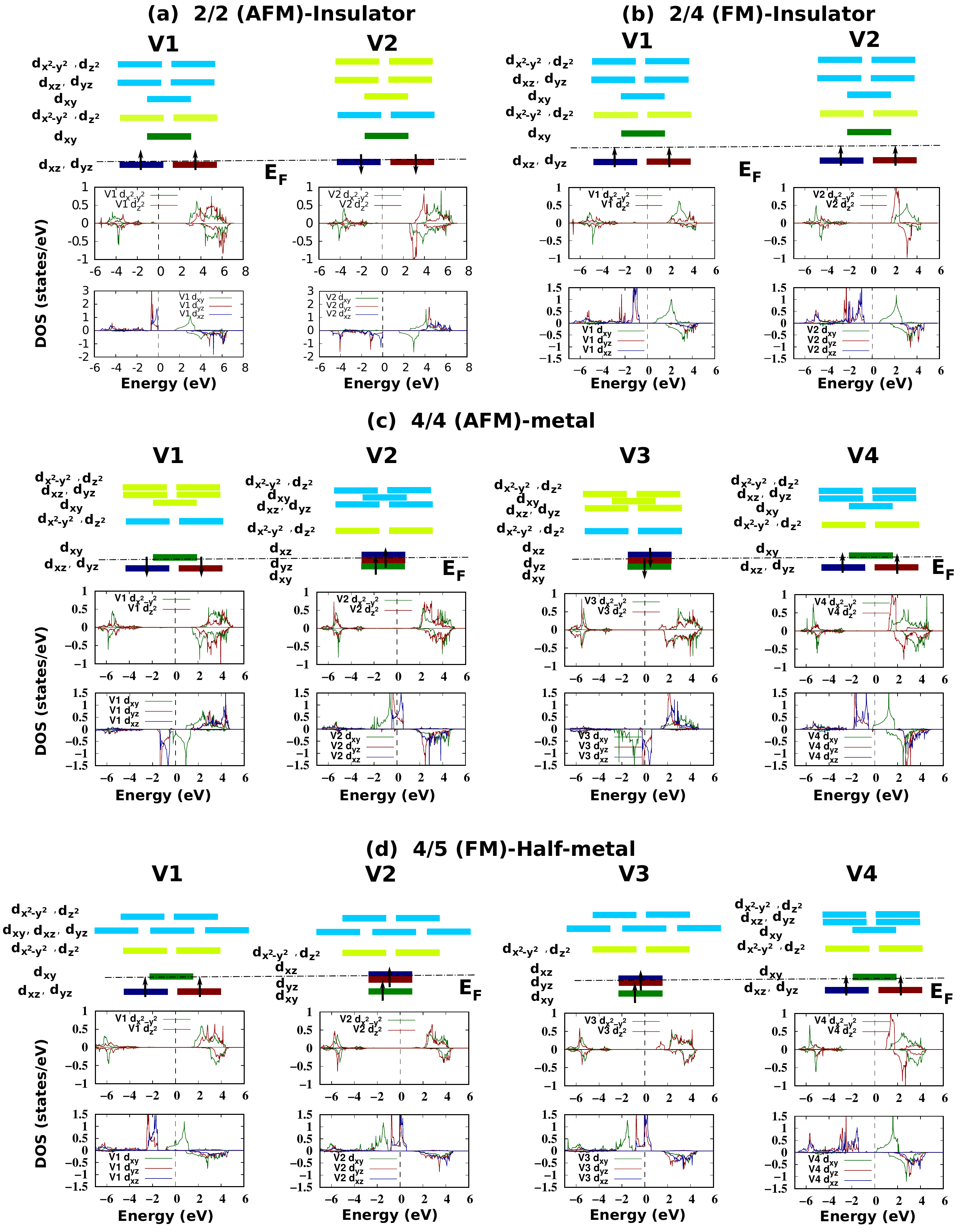}
		\caption{The orbital occupations with spin-levels shown in light green (spin-up) and cyan (spin-down) for (a) 2/2, (b) 2/4, (c) 4/4, and (d) 4/5 interfaces. V1, V2, etc., are the Vanadium ions (V$^{3+}$). The Vanadium ion nearest to the interface is V1. Black dashed lines are the Fermi levels in all cases. Corresponding orbital-projected spin-resolved DOS are shown in the lower panels: $e_g$ and $t_{2g}$ orbitals in the middle and lower panels, respectively.}
		\label{fig:sp-levels}
	\end{figure*}
	
	In order to understand the underlying mechanism for the magnetic ordering, we plot the orbital-levels and corresponding orbital-projected, spin-resolved density of states in Fig. \ref{fig:sp-levels} of different interfaces with different magnetic and conducting characters. Also shown are the primary location of the two electrons of V$^{3+}$ in its outermost shell. Analyzing the spin-resolved DOS as presented in Fig. \ref{fig:sp-dos}, we identified different magnetic characteristics for configurations 2/2, 2/4, 4/4, and 4/5, classifying them as FM-insulator, FM-\textit{half-metal}, AFM-metal, and FM-\textit{half-metal}, respectively. We report for all possible conducting and magnetic orderings for the sake of completeness, even though the interlayer AFM is the most stable phase. The Goodenough-Kanamori-Anderson (GKA) rule \cite{anderson1950antiferromagnetism,goodenough1954theory,goodenough1955theory,kanamori1959superexchange,wang2022goodenough} may be used to interpret the magnetic behavior of the 2/2 configuration shown in Fig. \ref{fig:sp-levels}(a). In the 2/2 heterostructure, antiferromagnetic coupling along the $c-$axis arises from a superexchange mechanism facilitated by oxygen anions. This can be seen from the DOS plot shown in Fig. S6 of SM \cite{patel2023layer}, where oxygen and V atom-projected DOS are present on the same energies just below the FL. This involves a virtual hop between the two V$^{3+}$ ions via the intervening oxygen. Specifically, the superexchange involves a $p-d$ hopping process where O $2p$ orbitals overlap with V $3d_{xz,yz}$ orbitals. Moreover, the $\sigma$ overlap of $2p$ orbitals with the $e_g$ orbitals is stronger than the $\pi$ overlap with the $t_{2g}$ orbitals, indicating a relatively weak antiferromagnetic superexchange in the case of the 2/2 interface. The DOS in Fig. \ref{fig:sp-levels}(b) is essential to understand the ferromagnetic exchange in 2/4. Here, both V1 and V2 exhibit the same spin-orbital order, featuring a gap where the FL is elevated compared to 2/2. The FM superexchange mechanism governs the behavior of the 2/4 FM insulator, wherein an occupied and an empty orbital exchange electrons via oxygen. This superexchange is relatively weak in the 2/4 scenario, as the $t_{2g}$ orbitals bond with O $2p$ orbitals through $\pi$ overlap.
		
	Next we analyze the DOS for the 4/4 interface where insulator-metal transition occurs (Shown in Fig. \ref{fig:sp-levels}(c)). Clearly, $t_{2g}$ orbitals are dominantly taking part at the FL. Focusing on a particular orbital, such as $d_{xy}$ (indicated by the green lines in the lower panels of Fig. \ref{fig:sp-levels}(c)), of V1 and V2, they touch each other just below the FL with their opposite spin moments. The same is true for V3 and V4 as shown in the last two panels of Fig. \ref{fig:sp-levels}(c). In this scenario, electrons hop between
	nearest V$^{3+}$ ions along the $c-$axis via intervening oxygen leading to interplanar antiferromagnetic superechange. A direct exchange mechanism can be utilized to explain the FM \textit{half-metal} character in 4/5, with no contribution from down spins from any of the V atoms. It is important to remark that in the 4/4 and 4/5 configurations of Fig. \ref{fig:sp-levels}(c-d), the inner V$^{3+}$ ions (V2 and V3) are alike in terms of DOS, while the outermost V$^{3+}$ ion (V4) is similar to the one near the interface (V1). This suggests there is an interaction which connects interface to the surface, such as electron reconstruction mechanism.

\begin{figure}[!htb]
		\centering
		\includegraphics[scale=0.35]{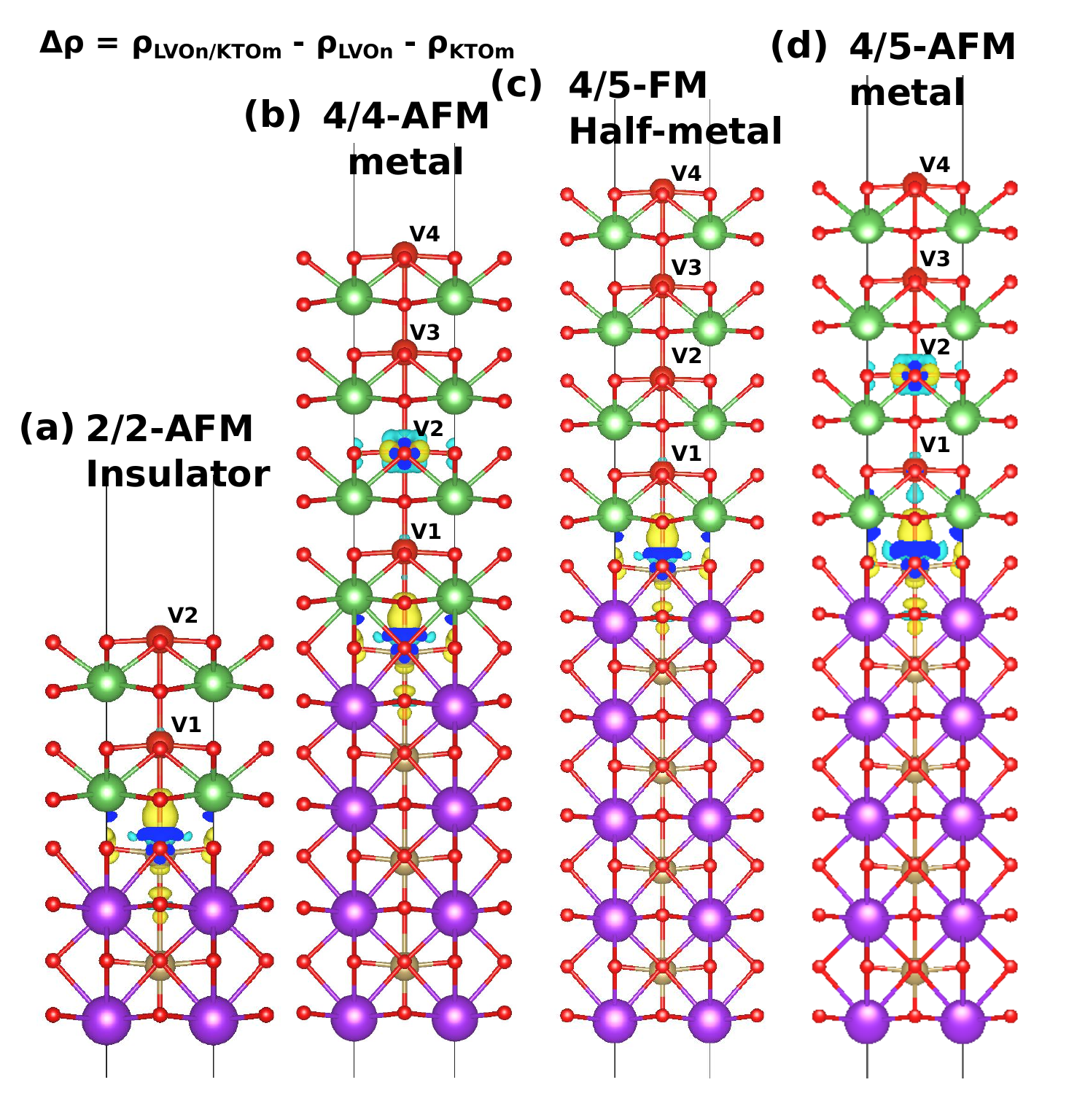}
		\caption{The charge density difference ($\Delta\rho$) plots for (a) 2/2, AFM-insulator, (b) 4/4, AFM-metal, (c) 4/5, FM-\textit{half-metal}, and (d) 4/5, AFM-metal.  Yellow surfaces represent electron clouds and the blue regions represent electron depletion. The isosurface value is set at $0.005\, e/$\AA$^3$.}
		\label{fig:cd_diff}
	\end{figure}

	The AFM-metallic phases are always intriguing. In order to get insights how AFM-metallic phases of 4/$n$ interfaces are different from AFM-insulator (2/2) and FM-\textit{half-metals}, and understanding the charge reconstruction mechanism, we plot the charge density difference in Fig. \ref{fig:cd_diff}. One thing which is common in all the four cases is the formation of two-dimensional electron gas (2DEG) at the interface. This indicates charge accumulation at the  interfacial $\rm{Ta_{int}}$, as will be further demonstrated by the partial free charge carriers we will discuss shortly. The formation of the interface attracts charges from the surface and bulk layers and induce electronic interactions at the interface of both the materials leading to conduction of the interface. The only difference among all is observed in the 4/$n$ cases (see Fig. \ref{fig:cd_diff}(b,d)), where it is clearly seen that there are more charge accumulation at the V2 site. As we mentioned earlier, 4/$n$ is the case where the AFM interfaces start conducting and showing a transition from insulating to metallic phase. We see that there is a charge leftover at V2 after subtracting the charge densities of LVO and KTO, and this leftover charge is supposed to be the source of the metallic AFM phase in the 4/$n$ interfaces. We also calculate partial free charge carrier stemming from different V atoms by integrating the atom-projected DOS near the FL for the 4/4 interface (Shown in Fig. S9 of SM \cite{patel2023layer}). There are mixed charge carriers (electrons and holes) at the interface whose magnitudes are given in Table T3 of SM \cite{patel2023layer}. We find that holes are in majority, which are contributed by the internal V atoms, V2 and V3. There is a small contribution of electrons at $\rm{Ta_{int}}$ (see Table T3 of SM \cite{patel2023layer}).
	
	We will now discuss the electronic polarization originated from the polar distortions in the insulating interfaces using Berry phase method \cite{king1993theory,stengel2009berry,stengel2011first}. Our focus lies specifically on the aggregate electronic polarization within magnetic insulating interfaces. A comprehensive examination of local polarizations and their associated distortions necessitates a separate, detailed investigation. The total electronic polarization calculated for 2/2-AFM, 2/3-AFM, and 3/3-AFM insulating interfaces are 0.24 $C/m^2$, 0.71 $C/m^2$, and 0.73 $C/m^2$, respectively. It is crucial to underscore that the relaxation of crystal geometries induces local distortions, resulting in a charge imbalance and the formation of dipoles. These dipoles constitute the primary origin of electronic polarization within the system. In contrast, the unrelaxed geometries exhibit have higher energy, are matallic, and do not support dipole moments.
	
	In the oxide interfaces of this nature, as highlighted in our introduction, achieving a critical thickness is pivotal for metallicity of the interface. Contrary to experimental claims \cite{wadehra2020planar} our study emphasizes the consideration of magnetic interactions for the determination of the critical thickness and there is no unique value for every $m/n$ combination. Our spin-polarized calculations demonstrate that, for the interface to exhibit conductivity, a critical thickness of no less than four layers of LVO, featuring interplanar AFM interactions, is imperative (Fig. \ref{fig:sp-dos}).

	\section{\label{sec:sqrt2} $\sqrt{2}\times\sqrt{2}$ supercells}
\begin{figure}[!ht]
		\centering
		\includegraphics[scale=0.6]{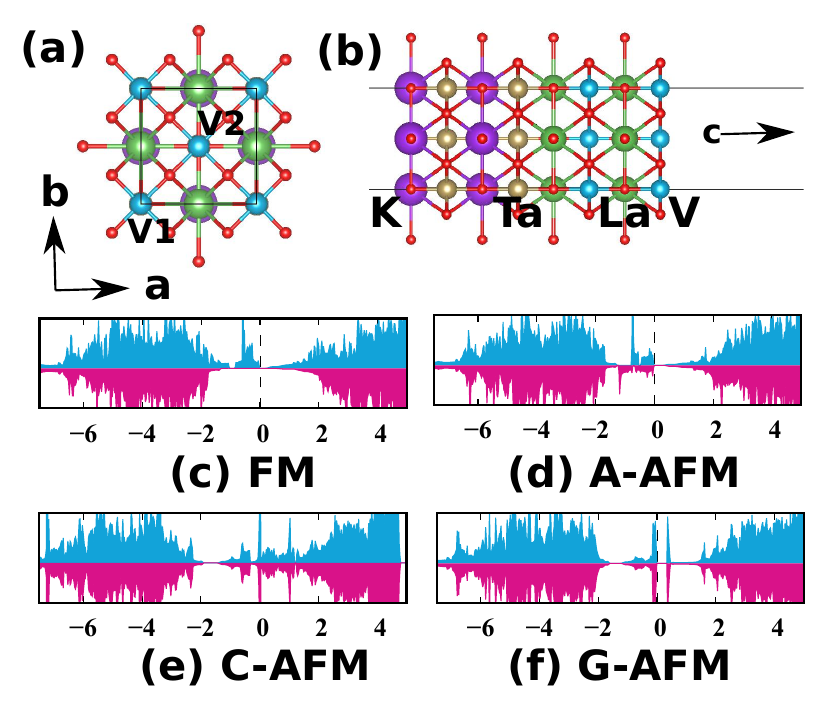}
		\caption{ (a) Top and (b) side-view of a $\sqrt{2}\times\sqrt{2}$ supercell for the 2/2 configuration. V1 and V2 indicate the magnetic V$^{3+}$ ions at the corner and at the center of the supercell, respectively. There are four magnetic ions in this 2/2 supercell, two in each layer, posing a possibility for FM, A-AFM, C-AFM, and G-AFM. Respective spin-polarized DOSs are shown in (c) FM, (d) A-AFM, (e) C-AFM, and (f) G-AFM. Vacuum is added along the $c-$direction as shown by the arrow in (b).}
		\label{fig:sqrt22}
	\end{figure}

	To explore the intra and interlayer magnetic coupling further and to confirm that intra-planar exchange between V moments is ferromagnetic, we construct a $\sqrt{2}\times\sqrt{2}$ supercell for the 2/2, 3/3 and 4/4 supercells. We expand the 1$\times$1 supercells along the diagonal direction, the in-plane lattice constant of the new supercells is now $\sqrt{2}a = 5.70$ \AA\, ($a$ is the lattice constant of 1$\times$1 supercell). Therefore, the new supercells have two transition-metal (TM) ions within each plane. In the $\sqrt{2}\times\sqrt{2}$ supercells, there are two magnetic V$^{3+}$ ions in each plane, one at the corner (V1) and another at the center (V2), the schematic of which is shown in Fig. \ref{fig:sqrt22}(a). A relaxed structure is also shown in Fig. S8 of SM \cite{patel2023layer} for the 3/3 heterostructure. All the atomic positions and the lattice geometries are allowed to relax and buckling in the bonds can be seen after the relaxation. The La-O and V-O (K-O and Ta-O) bonds bend in the downward (upward) direction along the $c$ axis and the overall bond angle (O-$M$-O, where $M$ is $A$ or $B$ of the $AB$O$_3$ complex) in the plane deviates from 180$^\circ$. As dicussed for the normal supercells, the $\sqrt{2}\times\sqrt{2}$ supercells also show an overall elongation along the $c$ axis. LVO and KTO layers stretch in the opposite directions inducing a polar distortion bewteen LVO and KTO (see Fig. S8(b) of SM \cite{patel2023layer}). The $\sqrt{2}\times\sqrt{2}$ supercells make it possible to study various AFM orders, such as A-AFM, C-AFM and G-AFM. In the A-AFM, all the magnetic ions in a plane are aligned along the z-direction while the ones in the next plane are oppositely aligned. In C-AFM order, the spins of the corner ions are up in each layer while the spins at the central atoms are down. In G-AFM, all the nearest-neighbor spins are aligned opposite to each other.

	\begin{center}
		\begin{table*}[!htb]
			\centering
			\setlength{\tabcolsep}{18pt}
			\renewcommand{\arraystretch}{1.2}
			\caption{\label{table:sqrt22} The ground state energies per unit cell (in eV) for  ferromagnetic (FM), A-type Antiferromagnetic (A-AFM), C-AFM, and G-AFM orderings for $\sqrt{2}\times\sqrt{2}$ supercells. The GSEs suggest that the A-AFM is the most stable and the FM is the second most stable state, except 1/n.}
			\begin{tabular}{ c c c c c c}
				\hline
				\hline
				interface &	FM & A-AFM & C-AFM & G-AFM & Nature\\
				\hline
				1/2	& -227.326 &  & -227.413 & & AFM-Insulator\\
				1/4	& -375.194 &  & -372.613 & & FM-Insulator\\
				2/2	& -307.876 & -307.985 & -305.313 & -306.408 & A-AFM-Insulator\\
				3/3	& -462.285 & -462.462 & -459.698 & -461.020 & A-AFM-Insulator\\
				4/4 & -616.250 & -616.591 & -613.942 & -616.337 & A-AFM-Metal\\
				\hline
			\end{tabular}
		\end{table*}
	\end{center}
	
	Comparing the GSEs for the normal $1\times 1$ and the $\sqrt{2}\times\sqrt{2}$ supercells, for FM and A-AFM order, we note that the GSEs are exactly doubled due to doubling of the system size, within the accuracy of $\sim$0.01 eV. We next calculate the GSEs for all the four (FM, A-AFM, C-AFM and G-AFM) magnetic orders, and find that the A-AFM is the most stable state and the FM is the second most stable,  except 1/2 and 1/4-$\sqrt{2}\times\sqrt{2}$ supercells. In 1/2 the intralayer AFM is more stable as compared to intralayer FM. The intralayer AFM in 1/2 is slightly lower in energy ($0.08$ eV) as compared to the FM. On the other hand, the trend in 1/4 is consistent with other m/m configuations (shown in Table \ref{table:sqrt22}) for which the intralayer AFM is the less stable as compared to the intralayer FM. The C-AFM magnetic state possesses highest GSE, implying that 2D antiferromagnetism (in a single plane) is unstable. Our study thus suggests that the interlayer AFM coupling (A-AFM) is favorable and the intralayer coupling is ferromagnetic (as against C and G-type AFM). Therefore, the stable AFM orders are, in fact, A-type AFM ground states. We plot the corresponding DOS in Fig. \ref{fig:sqrt22}(c-f). We notice that the FM and A-AFM are insulating with the valence bands almost touching but not crossing the FL. We also show that the DOS for the normal 2/2 supercell are insulating for the FM and the A-AFM states (Fig. \ref{fig:sp-dos}). The DOS for 1/2, 1/4, 3/3 and 4/4-$\sqrt{2}\times\sqrt{2}$ supercells are shown in Fig. S7 of the SM \cite{patel2023layer}. 1/2 and 1/4 are insulating in their stable phases, intralayer AFM and FM, respectively. For 3/3 and 4/4, the A-type AFM is the most stable state with insulating and metallic property, respectively, as shown in Figs. S7(i) and S7(n). Thus, we find that the critical thickness for metallicity of LVO/KTO heterostructures is more than three layers of LVO, that holds true even for the extended $\sqrt{2}\times\sqrt{2}$ supercells. In addition, magnetism plays an important role in confirming the experimentally observed insulator to metal transition.
	
	\section{\label{sec:conclusion}Discussion and Conclusions} 
	In summary, we studied several heterostructures formed out of cubic LVO and KTO perovskites, stacked along the (001) direction. We assessed their stability by computing their formation energies, with negative values indicating the feasibility of interfaces with cubic KTO as the substrate, despite LVO naturally existing in the orthorhombic phase. We study both non-magnetic and magnetic cases. Our findings for the non-magnetic case are as follows: (1) To have Ta $5d_{xy}$ states participate in transport, a minimum of two cubic LVO layers is necessary. (2) Mostly, $t_{2g}$ bands of all the $3d$-V atoms contribute to the FL, while only Ta$_{\rm{int}}\,5d_{xy}$ band plays any role. (3) The Lifshitz transition, observed in interfaces with three or more layers of LVO, involves V $3d_{xz,yz}$ bands near the interface and Ta $5d_{xy}$ at the interface, in contrast with LAO/STO interface, where the transition occurs among the $t_{2g}$ bands of Ti at the interface. This underscores the intricate multi-atom multi-orbital contributions to the transport phenomenon in these heterostructures. Furthermore, our study demonstrates that the bands involved in the Lifshitz transition can be finely tuned using the thickness of LVO and KTO. (4) Notably, our band structure and the model Hamiltonian study predicts negligible Rashba SOC, raising questions about the significance of SOC in transport. Additional layers of KTO, while keeping LVO layers constant, have minimal effects on the physics near the FL.
	
	Magnetism in LVO/KTO heterostructures, is primarily influenced by the magnetic properties of LVO. Spin-polarized DFT calculations indicate that in LVO/KTO heterostructures, the interlayer AFM ordering, commonly known as A-AFM, ground state is the most favorable, while the non-magnetic ground state being the least stable. We observe that if the number of LVO layers are fewer than four, the A-AFM ground states become insulating. The DOS for ferromagnetic heterostructures demonstrates a \textit{half-metallic} character. The small energy difference between FM and AFM ground state energies implies a potential coexistence of AFM and FM ground states. Superexchange interaction mediated via oxygen ions is responsible for the antiferromagnetic coupling, while a direct exchange most likely leads to ferromagnetic coupling. The magnetic interactions are necessary to explain the critical thickness of LVO for the conducting LVO/KTO heterostructures. The extended $\sqrt{2}\times\sqrt{2}$ supercells further confirms the essential role of magnetic interactions in elucidating the critical thickness of LVO. Furthermore, it is also shown that A-AFM (interlayer AFM and intralayer FM) is most stable state for all $m/n$, except 1/2, for which the intralayer AFM is favorable.
	
	\section{Acknowledgements}
	We acknowledge National Supercomputing Mission (NSM) for providing computing resources of 'PARAM Shakti' at IIT Kharagpur, which is implemented by C-DAC and supported by the Ministry of Electronics and Information Technology (MeitY) and Department of Science and Technology (DST), Government of India. The work at Los Alamos National Laboratory was carried out under the auspices of the U.S. Department of Energy (DOE) National Nuclear Security Administration under Contract No. 89233218CNA000001. It was supported by the LANL LDRD Program, and in part by the Center for Integrated Nanotechnologies, a DOE BES user facility, in partnership with the LANL Institutional Computing Program for computational resources.

	\bibliography{references}

\begin{thebibliography}{65}%
\makeatletter
\providecommand \@ifxundefined [1]{%
 \@ifx{#1\undefined}
}%
\providecommand \@ifnum [1]{%
 \ifnum #1\expandafter \@firstoftwo
 \else \expandafter \@secondoftwo
 \fi
}%
\providecommand \@ifx [1]{%
 \ifx #1\expandafter \@firstoftwo
 \else \expandafter \@secondoftwo
 \fi
}%
\providecommand \natexlab [1]{#1}%
\providecommand \enquote  [1]{``#1''}%
\providecommand \bibnamefont  [1]{#1}%
\providecommand \bibfnamefont [1]{#1}%
\providecommand \citenamefont [1]{#1}%
\providecommand \href@noop [0]{\@secondoftwo}%
\providecommand \href [0]{\begingroup \@sanitize@url \@href}%
\providecommand \@href[1]{\@@startlink{#1}\@@href}%
\providecommand \@@href[1]{\endgroup#1\@@endlink}%
\providecommand \@sanitize@url [0]{\catcode `\\12\catcode `\$12\catcode
  `\&12\catcode `\#12\catcode `\^12\catcode `\_12\catcode `\%12\relax}%
\providecommand \@@startlink[1]{}%
\providecommand \@@endlink[0]{}%
\providecommand \url  [0]{\begingroup\@sanitize@url \@url }%
\providecommand \@url [1]{\endgroup\@href {#1}{\urlprefix }}%
\providecommand \urlprefix  [0]{URL }%
\providecommand \Eprint [0]{\href }%
\providecommand \doibase [0]{https://doi.org/}%
\providecommand \selectlanguage [0]{\@gobble}%
\providecommand \bibinfo  [0]{\@secondoftwo}%
\providecommand \bibfield  [0]{\@secondoftwo}%
\providecommand \translation [1]{[#1]}%
\providecommand \BibitemOpen [0]{}%
\providecommand \bibitemStop [0]{}%
\providecommand \bibitemNoStop [0]{.\EOS\space}%
\providecommand \EOS [0]{\spacefactor3000\relax}%
\providecommand \BibitemShut  [1]{\csname bibitem#1\endcsname}%
\let\auto@bib@innerbib\@empty
\bibitem [{\citenamefont {Ohtomo}\ and\ \citenamefont
  {Hwang}(2004)}]{ohtomo2004high}%
  \BibitemOpen
  \bibfield  {author} {\bibinfo {author} {\bibfnamefont {A.}~\bibnamefont
  {Ohtomo}}\ and\ \bibinfo {author} {\bibfnamefont {H.}~\bibnamefont {Hwang}},\
  }\bibfield  {title} {\bibinfo {title} {A high-mobility electron gas at the
  {LaAlO$_3$/SrTiO$_3$} heterointerface},\ }\href
  {https://doi.org/10.1038/nature02308} {\bibfield  {journal} {\bibinfo
  {journal} {Nature}\ }\textbf {\bibinfo {volume} {427}},\ \bibinfo {pages}
  {423} (\bibinfo {year} {2004})}\BibitemShut {NoStop}%
\bibitem [{\citenamefont {Caviglia}\ \emph {et~al.}(2010)\citenamefont
  {Caviglia}, \citenamefont {Gariglio}, \citenamefont {Cancellieri},
  \citenamefont {ac{\'e}p{\'e}}, \citenamefont {Fete}, \citenamefont {Reyren},
  \citenamefont {Gabay}, \citenamefont {Morpurgo},\ and\ \citenamefont
  {Triscone}}]{caviglia2010two}%
  \BibitemOpen
  \bibfield  {author} {\bibinfo {author} {\bibfnamefont {A.~D.}\ \bibnamefont
  {Caviglia}}, \bibinfo {author} {\bibfnamefont {S.}~\bibnamefont {Gariglio}},
  \bibinfo {author} {\bibfnamefont {C.}~\bibnamefont {Cancellieri}}, \bibinfo
  {author} {\bibfnamefont {B.}~\bibnamefont {ac{\'e}p{\'e}}}, \bibinfo {author}
  {\bibfnamefont {A.}~\bibnamefont {Fete}}, \bibinfo {author} {\bibfnamefont
  {N.}~\bibnamefont {Reyren}}, \bibinfo {author} {\bibfnamefont
  {M.}~\bibnamefont {Gabay}}, \bibinfo {author} {\bibfnamefont {A.~F.}\
  \bibnamefont {Morpurgo}},\ and\ \bibinfo {author} {\bibfnamefont {J.-M.}\
  \bibnamefont {Triscone}},\ }\bibfield  {title} {\bibinfo {title}
  {Two-dimensional quantum oscillations of the conductance at
  {${\mathrm{LaAlO}}_{3}/{\mathrm{SrTiO}}_{3}$} interfaces},\ }\href
  {https://doi.org/10.1103/PhysRevLett.105.236802} {\bibfield  {journal}
  {\bibinfo  {journal} {Physical Review Letters}\ }\textbf {\bibinfo {volume}
  {105}},\ \bibinfo {pages} {236802} (\bibinfo {year} {2010})}\BibitemShut
  {NoStop}%
\bibitem [{\citenamefont {Bert}\ \emph {et~al.}(2011)\citenamefont {Bert},
  \citenamefont {Kalisky}, \citenamefont {Bell}, \citenamefont {Kim},
  \citenamefont {Hikita}, \citenamefont {Hwang},\ and\ \citenamefont
  {Moler}}]{bert2011direct}%
  \BibitemOpen
  \bibfield  {author} {\bibinfo {author} {\bibfnamefont {J.~A.}\ \bibnamefont
  {Bert}}, \bibinfo {author} {\bibfnamefont {B.}~\bibnamefont {Kalisky}},
  \bibinfo {author} {\bibfnamefont {C.}~\bibnamefont {Bell}}, \bibinfo {author}
  {\bibfnamefont {M.}~\bibnamefont {Kim}}, \bibinfo {author} {\bibfnamefont
  {Y.}~\bibnamefont {Hikita}}, \bibinfo {author} {\bibfnamefont {H.~Y.}\
  \bibnamefont {Hwang}},\ and\ \bibinfo {author} {\bibfnamefont {K.~A.}\
  \bibnamefont {Moler}},\ }\bibfield  {title} {\bibinfo {title} {Direct imaging
  of the coexistence of ferromagnetism and superconductivity at the
  {LaAlO$_3$/SrTiO$_3$} interface},\ }\href {https://doi.org/10.1038/nphys2079}
  {\bibfield  {journal} {\bibinfo  {journal} {Nature Physics}\ }\textbf
  {\bibinfo {volume} {7}},\ \bibinfo {pages} {767} (\bibinfo {year}
  {2011})}\BibitemShut {NoStop}%
\bibitem [{\citenamefont {Wadehra}\ \emph {et~al.}(2020)\citenamefont
  {Wadehra}, \citenamefont {Tomar}, \citenamefont {Varma}, \citenamefont
  {Gopal}, \citenamefont {Singh}, \citenamefont {Dattagupta},\ and\
  \citenamefont {Chakraverty}}]{wadehra2020planar}%
  \BibitemOpen
  \bibfield  {author} {\bibinfo {author} {\bibfnamefont {N.}~\bibnamefont
  {Wadehra}}, \bibinfo {author} {\bibfnamefont {R.}~\bibnamefont {Tomar}},
  \bibinfo {author} {\bibfnamefont {R.~M.}\ \bibnamefont {Varma}}, \bibinfo
  {author} {\bibfnamefont {R.}~\bibnamefont {Gopal}}, \bibinfo {author}
  {\bibfnamefont {Y.}~\bibnamefont {Singh}}, \bibinfo {author} {\bibfnamefont
  {S.}~\bibnamefont {Dattagupta}},\ and\ \bibinfo {author} {\bibfnamefont
  {S.}~\bibnamefont {Chakraverty}},\ }\bibfield  {title} {\bibinfo {title}
  {{Planar Hall effect and anisotropic magnetoresistance in polar-polar
  interface of LaVO$_3$-KTaO$_3$ with strong spin-orbit coupling}},\ }\href
  {https://doi.org/10.1038/s41467-020-14689-z} {\bibfield  {journal} {\bibinfo
  {journal} {Nature Communications}\ }\textbf {\bibinfo {volume} {11}},\
  \bibinfo {pages} {874} (\bibinfo {year} {2020})}\BibitemShut {NoStop}%
\bibitem [{\citenamefont {Wang}\ \emph {et~al.}(2016)\citenamefont {Wang},
  \citenamefont {Tang}, \citenamefont {Cheng}, \citenamefont {Behtash},\ and\
  \citenamefont {Yang}}]{wang2016creating}%
  \BibitemOpen
  \bibfield  {author} {\bibinfo {author} {\bibfnamefont {Y.}~\bibnamefont
  {Wang}}, \bibinfo {author} {\bibfnamefont {W.}~\bibnamefont {Tang}}, \bibinfo
  {author} {\bibfnamefont {J.}~\bibnamefont {Cheng}}, \bibinfo {author}
  {\bibfnamefont {M.}~\bibnamefont {Behtash}},\ and\ \bibinfo {author}
  {\bibfnamefont {K.}~\bibnamefont {Yang}},\ }\bibfield  {title} {\bibinfo
  {title} {{Creating two-dimensional electron gas in polar/polar perovskite
  oxide heterostructures: first-principles characterization of
  LaAlO$_3$/A$^+$B$^{5+}$O$_3$}},\ }\href
  {https://doi.org/10.1021/acsami.6b02399} {\bibfield  {journal} {\bibinfo
  {journal} {ACS applied materials \& interfaces}\ }\textbf {\bibinfo {volume}
  {8}},\ \bibinfo {pages} {13659} (\bibinfo {year} {2016})}\BibitemShut
  {NoStop}%
\bibitem [{\citenamefont {Thiel}\ \emph {et~al.}(2006)\citenamefont {Thiel},
  \citenamefont {Hammerl}, \citenamefont {Schmehl}, \citenamefont {Schneider},\
  and\ \citenamefont {Mannhart}}]{thiel2006tunable}%
  \BibitemOpen
  \bibfield  {author} {\bibinfo {author} {\bibfnamefont {S.}~\bibnamefont
  {Thiel}}, \bibinfo {author} {\bibfnamefont {G.}~\bibnamefont {Hammerl}},
  \bibinfo {author} {\bibfnamefont {A.}~\bibnamefont {Schmehl}}, \bibinfo
  {author} {\bibfnamefont {C.~W.}\ \bibnamefont {Schneider}},\ and\ \bibinfo
  {author} {\bibfnamefont {J.}~\bibnamefont {Mannhart}},\ }\bibfield  {title}
  {\bibinfo {title} {Tunable quasi-two-dimensional electron gases in oxide
  heterostructures},\ }\href {https://doi.org/10.1126/science.113109}
  {\bibfield  {journal} {\bibinfo  {journal} {Science}\ }\textbf {\bibinfo
  {volume} {313}},\ \bibinfo {pages} {1942} (\bibinfo {year}
  {2006})}\BibitemShut {NoStop}%
\bibitem [{\citenamefont {Pentcheva}\ and\ \citenamefont
  {Pickett}(2009)}]{pentcheva2009avoiding}%
  \BibitemOpen
  \bibfield  {author} {\bibinfo {author} {\bibfnamefont {R.}~\bibnamefont
  {Pentcheva}}\ and\ \bibinfo {author} {\bibfnamefont {W.~E.}\ \bibnamefont
  {Pickett}},\ }\bibfield  {title} {\bibinfo {title} {{Avoiding the
  Polarization Catastrophe in ${\mathrm{LaAlO}}_{3}$ Overlayers on
  ${\mathrm{SrTiO}}_{3}(001)$ through Polar Distortion}},\ }\href
  {https://doi.org/10.1103/PhysRevLett.102.107602} {\bibfield  {journal}
  {\bibinfo  {journal} {Physical Review Letters}\ }\textbf {\bibinfo {volume}
  {102}},\ \bibinfo {pages} {107602} (\bibinfo {year} {2009})}\BibitemShut
  {NoStop}%
\bibitem [{\citenamefont {Zhou}\ \emph {et~al.}(2015)\citenamefont {Zhou},
  \citenamefont {Asmara}, \citenamefont {Yang}, \citenamefont {Sawatzky},
  \citenamefont {Feng},\ and\ \citenamefont {Rusydi}}]{zhou2015interplay}%
  \BibitemOpen
  \bibfield  {author} {\bibinfo {author} {\bibfnamefont {J.}~\bibnamefont
  {Zhou}}, \bibinfo {author} {\bibfnamefont {T.~C.}\ \bibnamefont {Asmara}},
  \bibinfo {author} {\bibfnamefont {M.}~\bibnamefont {Yang}}, \bibinfo {author}
  {\bibfnamefont {G.~A.}\ \bibnamefont {Sawatzky}}, \bibinfo {author}
  {\bibfnamefont {Y.~P.}\ \bibnamefont {Feng}},\ and\ \bibinfo {author}
  {\bibfnamefont {A.}~\bibnamefont {Rusydi}},\ }\bibfield  {title} {\bibinfo
  {title} {{Interplay of electronic reconstructions, surface oxygen vacancies,
  and lattice distortions in insulator-metal transition of
  ${\mathrm{LaAlO}}_{3}/{\mathrm{SrTiO}}_{3}$}},\ }\href
  {https://doi.org/10.1103/PhysRevB.92.125423} {\bibfield  {journal} {\bibinfo
  {journal} {Physical Review B}\ }\textbf {\bibinfo {volume} {92}},\ \bibinfo
  {pages} {125423} (\bibinfo {year} {2015})}\BibitemShut {NoStop}%
\bibitem [{\citenamefont {Zhong}\ \emph {et~al.}(2013)\citenamefont {Zhong},
  \citenamefont {T{\'o}th},\ and\ \citenamefont {Held}}]{zhong2013theory}%
  \BibitemOpen
  \bibfield  {author} {\bibinfo {author} {\bibfnamefont {Z.}~\bibnamefont
  {Zhong}}, \bibinfo {author} {\bibfnamefont {A.}~\bibnamefont {T{\'o}th}},\
  and\ \bibinfo {author} {\bibfnamefont {K.}~\bibnamefont {Held}},\ }\bibfield
  {title} {\bibinfo {title} {Theory of spin-orbit coupling at
  {LaAlO$_3$/SrTiO$_3$} interfaces and {SrTiO$_3$} surfaces},\ }\href
  {https://doi.org/10.1103/PhysRevB.87.161102} {\bibfield  {journal} {\bibinfo
  {journal} {Physical Review B}\ }\textbf {\bibinfo {volume} {87}},\ \bibinfo
  {pages} {161102} (\bibinfo {year} {2013})}\BibitemShut {NoStop}%
\bibitem [{\citenamefont {Mohanta}\ and\ \citenamefont
  {Taraphder}(2015)}]{mohanta2015multiband}%
  \BibitemOpen
  \bibfield  {author} {\bibinfo {author} {\bibfnamefont {N.}~\bibnamefont
  {Mohanta}}\ and\ \bibinfo {author} {\bibfnamefont {A.}~\bibnamefont
  {Taraphder}},\ }\bibfield  {title} {\bibinfo {title} {{Multiband theory of
  superconductivity at the ${\mathrm{LaAlO}}_{3}/{\mathrm{SrTiO}}_{3}$
  interface}},\ }\href {https://doi.org/10.1103/PhysRevB.92.174531} {\bibfield
  {journal} {\bibinfo  {journal} {Physical Review B}\ }\textbf {\bibinfo
  {volume} {92}},\ \bibinfo {pages} {174531} (\bibinfo {year}
  {2015})}\BibitemShut {NoStop}%
\bibitem [{\citenamefont {Li}\ \emph {et~al.}(2011)\citenamefont {Li},
  \citenamefont {Richter}, \citenamefont {Mannhart},\ and\ \citenamefont
  {Ashoori}}]{li2011coexistence}%
  \BibitemOpen
  \bibfield  {author} {\bibinfo {author} {\bibfnamefont {L.}~\bibnamefont
  {Li}}, \bibinfo {author} {\bibfnamefont {C.}~\bibnamefont {Richter}},
  \bibinfo {author} {\bibfnamefont {J.}~\bibnamefont {Mannhart}},\ and\
  \bibinfo {author} {\bibfnamefont {R.}~\bibnamefont {Ashoori}},\ }\bibfield
  {title} {\bibinfo {title} {{Coexistence of magnetic order and two-dimensional
  superconductivity at LaAlO3/SrTiO3 interfaces}},\ }\href
  {https://doi.org/10.1038/nphys2080} {\bibfield  {journal} {\bibinfo
  {journal} {Nature physics}\ }\textbf {\bibinfo {volume} {7}},\ \bibinfo
  {pages} {762} (\bibinfo {year} {2011})}\BibitemShut {NoStop}%
\bibitem [{\citenamefont {Kalabukhov}\ \emph {et~al.}(2007)\citenamefont
  {Kalabukhov}, \citenamefont {Gunnarsson}, \citenamefont {B{\"o}rjesson},
  \citenamefont {Olsson}, \citenamefont {Claeson},\ and\ \citenamefont
  {Winkler}}]{kalabukhov2007effect}%
  \BibitemOpen
  \bibfield  {author} {\bibinfo {author} {\bibfnamefont {A.}~\bibnamefont
  {Kalabukhov}}, \bibinfo {author} {\bibfnamefont {R.}~\bibnamefont
  {Gunnarsson}}, \bibinfo {author} {\bibfnamefont {J.}~\bibnamefont
  {B{\"o}rjesson}}, \bibinfo {author} {\bibfnamefont {E.}~\bibnamefont
  {Olsson}}, \bibinfo {author} {\bibfnamefont {T.}~\bibnamefont {Claeson}},\
  and\ \bibinfo {author} {\bibfnamefont {D.}~\bibnamefont {Winkler}},\
  }\bibfield  {title} {\bibinfo {title} {{Effect of oxygen vacancies in the
  {SrTiO$_3$} substrate on the electrical properties of the
  {LaAlO$_3$/SrTiO$_3$} interface}},\ }\href
  {https://doi.org/10.1103/PhysRevB.75.121404} {\bibfield  {journal} {\bibinfo
  {journal} {Physical Review B}\ }\textbf {\bibinfo {volume} {75}},\ \bibinfo
  {pages} {121404} (\bibinfo {year} {2007})}\BibitemShut {NoStop}%
\bibitem [{\citenamefont {Mohanta}\ and\ \citenamefont
  {Taraphder}(2014)}]{mohanta2014topological}%
  \BibitemOpen
  \bibfield  {author} {\bibinfo {author} {\bibfnamefont {N.}~\bibnamefont
  {Mohanta}}\ and\ \bibinfo {author} {\bibfnamefont {A.}~\bibnamefont
  {Taraphder}},\ }\bibfield  {title} {\bibinfo {title} {Topological
  superconductivity and majorana bound states at the {LaAlO$_3$/SrTiO$_3$}
  interface},\ }\href {https://doi.org/10.1209/0295-5075/108/60001} {\bibfield
  {journal} {\bibinfo  {journal} {Europhysics Letters}\ }\textbf {\bibinfo
  {volume} {108}},\ \bibinfo {pages} {60001} (\bibinfo {year}
  {2014})}\BibitemShut {NoStop}%
\bibitem [{\citenamefont {Scheurer}\ and\ \citenamefont
  {Schmalian}(2015)}]{scheurer2015topological}%
  \BibitemOpen
  \bibfield  {author} {\bibinfo {author} {\bibfnamefont {M.~S.}\ \bibnamefont
  {Scheurer}}\ and\ \bibinfo {author} {\bibfnamefont {J.}~\bibnamefont
  {Schmalian}},\ }\bibfield  {title} {\bibinfo {title} {Topological
  superconductivity and unconventional pairing in oxide interfaces},\ }\href
  {https://doi.org/10.1038/ncomms7005} {\bibfield  {journal} {\bibinfo
  {journal} {Nature Communications}\ }\textbf {\bibinfo {volume} {6}},\
  \bibinfo {pages} {6005} (\bibinfo {year} {2015})}\BibitemShut {NoStop}%
\bibitem [{\citenamefont {Nandy}\ \emph {et~al.}(2016)\citenamefont {Nandy},
  \citenamefont {Mohanta}, \citenamefont {Acharya},\ and\ \citenamefont
  {Taraphder}}]{nandy2016anomalous}%
  \BibitemOpen
  \bibfield  {author} {\bibinfo {author} {\bibfnamefont {S.}~\bibnamefont
  {Nandy}}, \bibinfo {author} {\bibfnamefont {N.}~\bibnamefont {Mohanta}},
  \bibinfo {author} {\bibfnamefont {S.}~\bibnamefont {Acharya}},\ and\ \bibinfo
  {author} {\bibfnamefont {A.}~\bibnamefont {Taraphder}},\ }\bibfield  {title}
  {\bibinfo {title} {{Anomalous transport near the Lifshitz transition at the
  ${\mathrm{LaAlO}}_{3}/{\mathrm{SrTiO}}_{3}$ interface}},\ }\href
  {https://doi.org/10.1103/PhysRevB.94.155103} {\bibfield  {journal} {\bibinfo
  {journal} {Physical Review B}\ }\textbf {\bibinfo {volume} {94}},\ \bibinfo
  {pages} {155103} (\bibinfo {year} {2016})}\BibitemShut {NoStop}%
\bibitem [{\citenamefont {Popovi{\'c}}\ \emph {et~al.}(2008)\citenamefont
  {Popovi{\'c}}, \citenamefont {Satpathy},\ and\ \citenamefont
  {Martin}}]{popovic2008origin}%
  \BibitemOpen
  \bibfield  {author} {\bibinfo {author} {\bibfnamefont {Z.~S.}\ \bibnamefont
  {Popovi{\'c}}}, \bibinfo {author} {\bibfnamefont {S.}~\bibnamefont
  {Satpathy}},\ and\ \bibinfo {author} {\bibfnamefont {R.~M.}\ \bibnamefont
  {Martin}},\ }\bibfield  {title} {\bibinfo {title} {{Origin of the
  Two-Dimensional Electron Gas Carrier Density at the ${\mathrm{LaAlO}}_{3}$ on
  ${\mathrm{SrTiO}}_{3}$ Interface}},\ }\href
  {https://doi.org/10.1103/PhysRevLett.101.256801} {\bibfield  {journal}
  {\bibinfo  {journal} {Physical Review Letters}\ }\textbf {\bibinfo {volume}
  {101}},\ \bibinfo {pages} {256801} (\bibinfo {year} {2008})}\BibitemShut
  {NoStop}%
\bibitem [{\citenamefont {Nandy}\ \emph {et~al.}(2017)\citenamefont {Nandy},
  \citenamefont {Sharma}, \citenamefont {Taraphder},\ and\ \citenamefont
  {Tewari}}]{nandy2017chiral}%
  \BibitemOpen
  \bibfield  {author} {\bibinfo {author} {\bibfnamefont {S.}~\bibnamefont
  {Nandy}}, \bibinfo {author} {\bibfnamefont {G.}~\bibnamefont {Sharma}},
  \bibinfo {author} {\bibfnamefont {A.}~\bibnamefont {Taraphder}},\ and\
  \bibinfo {author} {\bibfnamefont {S.}~\bibnamefont {Tewari}},\ }\bibfield
  {title} {\bibinfo {title} {Chiral anomaly as the origin of the planar hall
  effect in weyl semimetals},\ }\href
  {https://doi.org/10.1103/PhysRevLett.119.176804} {\bibfield  {journal}
  {\bibinfo  {journal} {Physical Review Letters}\ }\textbf {\bibinfo {volume}
  {119}},\ \bibinfo {pages} {176804} (\bibinfo {year} {2017})}\BibitemShut
  {NoStop}%
\bibitem [{\citenamefont {Wemple}(1965)}]{wemple1965some}%
  \BibitemOpen
  \bibfield  {author} {\bibinfo {author} {\bibfnamefont {S.~H.}\ \bibnamefont
  {Wemple}},\ }\bibfield  {title} {\bibinfo {title} {{Some Transport Properties
  of Oxygen-Deficient Single-Crystal Potassium Tantalate
  (KTa${\mathrm{O}}_{3}$)}},\ }\href
  {https://doi.org/10.1103/PhysRev.137.A1575} {\bibfield  {journal} {\bibinfo
  {journal} {Physical Review}\ }\textbf {\bibinfo {volume} {137}},\ \bibinfo
  {pages} {A1575} (\bibinfo {year} {1965})}\BibitemShut {NoStop}%
\bibitem [{\citenamefont {Jellison}\ \emph {et~al.}(2006)\citenamefont
  {Jellison}, \citenamefont {Paulauskas}, \citenamefont {Boatner},\ and\
  \citenamefont {Singh}}]{jellison2006optical}%
  \BibitemOpen
  \bibfield  {author} {\bibinfo {author} {\bibfnamefont {G.~E.}\ \bibnamefont
  {Jellison}}, \bibinfo {author} {\bibfnamefont {I.}~\bibnamefont
  {Paulauskas}}, \bibinfo {author} {\bibfnamefont {L.~A.}\ \bibnamefont
  {Boatner}},\ and\ \bibinfo {author} {\bibfnamefont {D.~J.}\ \bibnamefont
  {Singh}},\ }\bibfield  {title} {\bibinfo {title} {{Optical functions of
  ${\mathrm{KTaO}}_{3}$ as determined by spectroscopic ellipsometry and
  comparison with band structure calculations}},\ }\href
  {https://doi.org/10.1103/PhysRevB.74.155130} {\bibfield  {journal} {\bibinfo
  {journal} {Physical Review B}\ }\textbf {\bibinfo {volume} {74}},\ \bibinfo
  {pages} {155130} (\bibinfo {year} {2006})}\BibitemShut {NoStop}%
\bibitem [{\citenamefont {Nakamura}\ and\ \citenamefont
  {Kimura}(2009)}]{nakamura2009electric}%
  \BibitemOpen
  \bibfield  {author} {\bibinfo {author} {\bibfnamefont {H.}~\bibnamefont
  {Nakamura}}\ and\ \bibinfo {author} {\bibfnamefont {T.}~\bibnamefont
  {Kimura}},\ }\bibfield  {title} {\bibinfo {title} {Electric field tuning of
  spin-orbit coupling in {${\text{KTaO}}_{3}$} field-effect transistors},\
  }\href {https://doi.org/10.1103/PhysRevB.80.121308} {\bibfield  {journal}
  {\bibinfo  {journal} {Physical Review B}\ }\textbf {\bibinfo {volume} {80}},\
  \bibinfo {pages} {121308} (\bibinfo {year} {2009})}\BibitemShut {NoStop}%
\bibitem [{\citenamefont {Shanavas}\ and\ \citenamefont
  {Satpathy}(2014)}]{shanavas2014electric}%
  \BibitemOpen
  \bibfield  {author} {\bibinfo {author} {\bibfnamefont {K.~V.}\ \bibnamefont
  {Shanavas}}\ and\ \bibinfo {author} {\bibfnamefont {S.}~\bibnamefont
  {Satpathy}},\ }\bibfield  {title} {\bibinfo {title} {Electric field tuning of
  the {Rashba} effect in the polar perovskite structures},\ }\href
  {https://doi.org/10.1103/PhysRevLett.112.086802} {\bibfield  {journal}
  {\bibinfo  {journal} {Physical Review Letters}\ }\textbf {\bibinfo {volume}
  {112}},\ \bibinfo {pages} {086802} (\bibinfo {year} {2014})}\BibitemShut
  {NoStop}%
\bibitem [{\citenamefont {Shanavas}(2015)}]{shanavas2015overview}%
  \BibitemOpen
  \bibfield  {author} {\bibinfo {author} {\bibfnamefont {K.}~\bibnamefont
  {Shanavas}},\ }\bibfield  {title} {\bibinfo {title} {Overview of theoretical
  studies of {Rashba} effect in polar perovskite surfaces},\ }\href
  {https://doi.org/10.1016/j.elspec.2014.08.005} {\bibfield  {journal}
  {\bibinfo  {journal} {Journal of Electron Spectroscopy and Related
  Phenomena}\ }\textbf {\bibinfo {volume} {201}},\ \bibinfo {pages} {121}
  (\bibinfo {year} {2015})}\BibitemShut {NoStop}%
\bibitem [{\citenamefont {Varignon}\ \emph {et~al.}(2019)\citenamefont
  {Varignon}, \citenamefont {Bibes},\ and\ \citenamefont
  {Zunger}}]{varignon2019origin}%
  \BibitemOpen
  \bibfield  {author} {\bibinfo {author} {\bibfnamefont {J.}~\bibnamefont
  {Varignon}}, \bibinfo {author} {\bibfnamefont {M.}~\bibnamefont {Bibes}},\
  and\ \bibinfo {author} {\bibfnamefont {A.}~\bibnamefont {Zunger}},\
  }\bibfield  {title} {\bibinfo {title} {Origin of band gaps in 3$d$ perovskite
  oxides},\ }\href {https://doi.org/10.1038/s41467-019-09698-6} {\bibfield
  {journal} {\bibinfo  {journal} {Nature communications}\ }\textbf {\bibinfo
  {volume} {10}},\ \bibinfo {pages} {1658} (\bibinfo {year}
  {2019})}\BibitemShut {NoStop}%
\bibitem [{\citenamefont {Malyi}\ \emph {et~al.}(2023)\citenamefont {Malyi},
  \citenamefont {Zhao},\ and\ \citenamefont {Zunger}}]{malyi2023insulating}%
  \BibitemOpen
  \bibfield  {author} {\bibinfo {author} {\bibfnamefont {O.~I.}\ \bibnamefont
  {Malyi}}, \bibinfo {author} {\bibfnamefont {X.-G.}\ \bibnamefont {Zhao}},\
  and\ \bibinfo {author} {\bibfnamefont {A.}~\bibnamefont {Zunger}},\
  }\bibfield  {title} {\bibinfo {title} {{Insulating band gaps both below and
  above the N\'eel temperature in $d$-electron $\mathrm{LaTi}{\mathrm{O}}_{3}$,
  $\mathrm{LaV}{\mathrm{O}}_{3}$, $\mathrm{SrMn}{\mathrm{O}}_{3}$, and
  $\mathrm{LaMn}{\mathrm{O}}_{3}$ perovskites as a symmetry-breaking event}},\
  }\href {https://doi.org/10.1103/PhysRevMaterials.7.074406} {\bibfield
  {journal} {\bibinfo  {journal} {Physical Review Materials}\ }\textbf
  {\bibinfo {volume} {7}},\ \bibinfo {pages} {074406} (\bibinfo {year}
  {2023})}\BibitemShut {NoStop}%
\bibitem [{\citenamefont {Kakkar}\ and\ \citenamefont
  {Bera}(2022)}]{kakkar2022rashba}%
  \BibitemOpen
  \bibfield  {author} {\bibinfo {author} {\bibfnamefont {S.}~\bibnamefont
  {Kakkar}}\ and\ \bibinfo {author} {\bibfnamefont {C.}~\bibnamefont {Bera}},\
  }\bibfield  {title} {\bibinfo {title} {{Rashba spin splitting in
  two-dimensional electron gas in polar-polar perovskite oxide heterostructure
  {LaVO$_3$/KTaO$_3$}: A DFT investigation}},\ }\href
  {https://doi.org/10.1016/j.physe.2022.115394} {\bibfield  {journal} {\bibinfo
   {journal} {Physica E: Low-dimensional Systems and Nanostructures}\ }\textbf
  {\bibinfo {volume} {144}},\ \bibinfo {pages} {115394} (\bibinfo {year}
  {2022})}\BibitemShut {NoStop}%
\bibitem [{\citenamefont {Wold}\ and\ \citenamefont
  {Ward}(1954)}]{wold1954perowskite}%
  \BibitemOpen
  \bibfield  {author} {\bibinfo {author} {\bibfnamefont {A.}~\bibnamefont
  {Wold}}\ and\ \bibinfo {author} {\bibfnamefont {R.}~\bibnamefont {Ward}},\
  }\bibfield  {title} {\bibinfo {title} {Perowskite-type oxides of cobalt,
  chromium and vanadium with some rare earth elements},\ }\href
  {https://doi.org/10.1021/ja01633a031} {\bibfield  {journal} {\bibinfo
  {journal} {Journal of the American Chemical Society}\ }\textbf {\bibinfo
  {volume} {76}},\ \bibinfo {pages} {1029} (\bibinfo {year}
  {1954})}\BibitemShut {NoStop}%
\bibitem [{\citenamefont {Jiang}\ \emph {et~al.}(2006)\citenamefont {Jiang},
  \citenamefont {Guo}, \citenamefont {Liu}, \citenamefont {Zhu}, \citenamefont
  {Zhou}, \citenamefont {Wu},\ and\ \citenamefont {Li}}]{jiang2006prediction}%
  \BibitemOpen
  \bibfield  {author} {\bibinfo {author} {\bibfnamefont {L.}~\bibnamefont
  {Jiang}}, \bibinfo {author} {\bibfnamefont {J.}~\bibnamefont {Guo}}, \bibinfo
  {author} {\bibfnamefont {H.}~\bibnamefont {Liu}}, \bibinfo {author}
  {\bibfnamefont {M.}~\bibnamefont {Zhu}}, \bibinfo {author} {\bibfnamefont
  {X.}~\bibnamefont {Zhou}}, \bibinfo {author} {\bibfnamefont {P.}~\bibnamefont
  {Wu}},\ and\ \bibinfo {author} {\bibfnamefont {C.}~\bibnamefont {Li}},\
  }\bibfield  {title} {\bibinfo {title} {Prediction of lattice constant in
  cubic perovskites},\ }\href {https://doi.org/10.1016/j.jpcs.2006.02.004}
  {\bibfield  {journal} {\bibinfo  {journal} {Journal of Physics and Chemistry
  of Solids}\ }\textbf {\bibinfo {volume} {67}},\ \bibinfo {pages} {1531}
  (\bibinfo {year} {2006})}\BibitemShut {NoStop}%
\bibitem [{\citenamefont {Bordet}\ \emph {et~al.}(1993)\citenamefont {Bordet},
  \citenamefont {Chaillout}, \citenamefont {Marezio}, \citenamefont {Huang},
  \citenamefont {Santoro}, \citenamefont {Cheong}, \citenamefont {Takagi},
  \citenamefont {Oglesby},\ and\ \citenamefont
  {Batlogg}}]{bordet1993structural}%
  \BibitemOpen
  \bibfield  {author} {\bibinfo {author} {\bibfnamefont {P.}~\bibnamefont
  {Bordet}}, \bibinfo {author} {\bibfnamefont {C.}~\bibnamefont {Chaillout}},
  \bibinfo {author} {\bibfnamefont {M.}~\bibnamefont {Marezio}}, \bibinfo
  {author} {\bibfnamefont {Q.}~\bibnamefont {Huang}}, \bibinfo {author}
  {\bibfnamefont {A.}~\bibnamefont {Santoro}}, \bibinfo {author} {\bibfnamefont
  {S.}~\bibnamefont {Cheong}}, \bibinfo {author} {\bibfnamefont
  {H.}~\bibnamefont {Takagi}}, \bibinfo {author} {\bibfnamefont
  {C.}~\bibnamefont {Oglesby}},\ and\ \bibinfo {author} {\bibfnamefont
  {B.}~\bibnamefont {Batlogg}},\ }\bibfield  {title} {\bibinfo {title}
  {Structural aspects of the crystallographic-magnetic transition in {LaVO$_3$}
  around 140 k},\ }\href {https://doi.org/10.1006/jssc.1993.1285} {\bibfield
  {journal} {\bibinfo  {journal} {Journal of Solid State Chemistry}\ }\textbf
  {\bibinfo {volume} {106}},\ \bibinfo {pages} {253} (\bibinfo {year}
  {1993})}\BibitemShut {NoStop}%
\bibitem [{\citenamefont {De~Raychaudhury}\ \emph {et~al.}(2007)\citenamefont
  {De~Raychaudhury}, \citenamefont {Pavarini},\ and\ \citenamefont
  {Andersen}}]{de2007orbital}%
  \BibitemOpen
  \bibfield  {author} {\bibinfo {author} {\bibfnamefont {M.}~\bibnamefont
  {De~Raychaudhury}}, \bibinfo {author} {\bibfnamefont {E.}~\bibnamefont
  {Pavarini}},\ and\ \bibinfo {author} {\bibfnamefont {O.~K.}\ \bibnamefont
  {Andersen}},\ }\bibfield  {title} {\bibinfo {title} {{Orbital Fluctuations in
  the Different Phases of ${\mathrm{LaVO}}_{3}$ and ${\mathrm{YVO}}_{3}$}},\
  }\href {https://doi.org/10.1103/PhysRevLett.99.126402} {\bibfield  {journal}
  {\bibinfo  {journal} {Physical Review Letters}\ }\textbf {\bibinfo {volume}
  {99}},\ \bibinfo {pages} {126402} (\bibinfo {year} {2007})}\BibitemShut
  {NoStop}%
\bibitem [{\citenamefont {Miyasaka}\ \emph {et~al.}(2000)\citenamefont
  {Miyasaka}, \citenamefont {Okuda},\ and\ \citenamefont
  {Tokura}}]{miyasaka2000critical}%
  \BibitemOpen
  \bibfield  {author} {\bibinfo {author} {\bibfnamefont {S.}~\bibnamefont
  {Miyasaka}}, \bibinfo {author} {\bibfnamefont {T.}~\bibnamefont {Okuda}},\
  and\ \bibinfo {author} {\bibfnamefont {Y.}~\bibnamefont {Tokura}},\
  }\bibfield  {title} {\bibinfo {title} {{Critical Behavior of Metal-Insulator
  Transition in La$_{1- x}$Sr$_x$VO$_3$}},\ }\href
  {https://doi.org/10.1103/PhysRevLett.85.5388} {\bibfield  {journal} {\bibinfo
   {journal} {Physical Review Letters}\ }\textbf {\bibinfo {volume} {85}},\
  \bibinfo {pages} {5388} (\bibinfo {year} {2000})}\BibitemShut {NoStop}%
\bibitem [{\citenamefont {Hotta}\ \emph {et~al.}(2007)\citenamefont {Hotta},
  \citenamefont {Susaki},\ and\ \citenamefont {Hwang}}]{hotta2007polar}%
  \BibitemOpen
  \bibfield  {author} {\bibinfo {author} {\bibfnamefont {Y.}~\bibnamefont
  {Hotta}}, \bibinfo {author} {\bibfnamefont {T.}~\bibnamefont {Susaki}},\ and\
  \bibinfo {author} {\bibfnamefont {H.~Y.}\ \bibnamefont {Hwang}},\ }\bibfield
  {title} {\bibinfo {title} {{Polar Discontinuity Doping of the
  ${\mathrm{LaVO}}_{3}/{\mathrm{SrTiO}}_{3}$ Interface}},\ }\href
  {https://doi.org/10.1103/PhysRevLett.99.236805} {\bibfield  {journal}
  {\bibinfo  {journal} {Physical Review Letters}\ }\textbf {\bibinfo {volume}
  {99}},\ \bibinfo {pages} {236805} (\bibinfo {year} {2007})}\BibitemShut
  {NoStop}%
\bibitem [{\citenamefont {Rashid}\ \emph {et~al.}(2017)\citenamefont {Rashid},
  \citenamefont {Abbas}, \citenamefont {Yaseen}, \citenamefont {Afzal},
  \citenamefont {Mahmood},\ and\ \citenamefont
  {Ramay}}]{rashid2017theoretical}%
  \BibitemOpen
  \bibfield  {author} {\bibinfo {author} {\bibfnamefont {M.}~\bibnamefont
  {Rashid}}, \bibinfo {author} {\bibfnamefont {Z.}~\bibnamefont {Abbas}},
  \bibinfo {author} {\bibfnamefont {M.}~\bibnamefont {Yaseen}}, \bibinfo
  {author} {\bibfnamefont {Q.}~\bibnamefont {Afzal}}, \bibinfo {author}
  {\bibfnamefont {A.}~\bibnamefont {Mahmood}},\ and\ \bibinfo {author}
  {\bibfnamefont {S.~M.}\ \bibnamefont {Ramay}},\ }\bibfield  {title} {\bibinfo
  {title} {{Theoretical Investigation of Cubic BaVO$_3$ and LaVO$_3$
  Perovskites via Tran--Blaha-Modified Becke--Johnson Exchange Potential
  Approach}},\ }\href {https://doi.org/10.1007/s10948-017-4099-0} {\bibfield
  {journal} {\bibinfo  {journal} {Journal of Superconductivity and Novel
  Magnetism}\ }\textbf {\bibinfo {volume} {30}},\ \bibinfo {pages} {3129}
  (\bibinfo {year} {2017})}\BibitemShut {NoStop}%
\bibitem [{\citenamefont {L{\"u}ders}\ \emph {et~al.}(2009)\citenamefont
  {L{\"u}ders}, \citenamefont {Sheets}, \citenamefont {David}, \citenamefont
  {Prellier},\ and\ \citenamefont {Fr{\'e}sard}}]{luders2009room}%
  \BibitemOpen
  \bibfield  {author} {\bibinfo {author} {\bibfnamefont {U.}~\bibnamefont
  {L{\"u}ders}}, \bibinfo {author} {\bibfnamefont {W.~C.}\ \bibnamefont
  {Sheets}}, \bibinfo {author} {\bibfnamefont {A.}~\bibnamefont {David}},
  \bibinfo {author} {\bibfnamefont {W.}~\bibnamefont {Prellier}},\ and\
  \bibinfo {author} {\bibfnamefont {R.}~\bibnamefont {Fr{\'e}sard}},\
  }\bibfield  {title} {\bibinfo {title} {{Room-temperature magnetism in
  ${\text{LaVO}}_{3}/{\text{SrVO}}_{3}$ superlattices by geometrically confined
  doping}},\ }\href {https://doi.org/10.1103/PhysRevB.80.241102} {\bibfield
  {journal} {\bibinfo  {journal} {Physical Review B}\ }\textbf {\bibinfo
  {volume} {80}},\ \bibinfo {pages} {241102} (\bibinfo {year}
  {2009})}\BibitemShut {NoStop}%
\bibitem [{\citenamefont {Dang}\ and\ \citenamefont
  {Millis}(2013{\natexlab{a}})}]{dang2013designing}%
  \BibitemOpen
  \bibfield  {author} {\bibinfo {author} {\bibfnamefont {H.~T.}\ \bibnamefont
  {Dang}}\ and\ \bibinfo {author} {\bibfnamefont {A.~J.}\ \bibnamefont
  {Millis}},\ }\bibfield  {title} {\bibinfo {title} {Designing ferromagnetism
  in vanadium oxide based superlattices},\ }\href
  {https://doi.org/10.1103/PhysRevB.87.184434} {\bibfield  {journal} {\bibinfo
  {journal} {Physical Review B}\ }\textbf {\bibinfo {volume} {87}},\ \bibinfo
  {pages} {184434} (\bibinfo {year} {2013}{\natexlab{a}})}\BibitemShut
  {NoStop}%
\bibitem [{\citenamefont {Dang}\ and\ \citenamefont
  {Millis}(2013{\natexlab{b}})}]{dang2013theory}%
  \BibitemOpen
  \bibfield  {author} {\bibinfo {author} {\bibfnamefont {H.~T.}\ \bibnamefont
  {Dang}}\ and\ \bibinfo {author} {\bibfnamefont {A.~J.}\ \bibnamefont
  {Millis}},\ }\bibfield  {title} {\bibinfo {title} {Theory of ferromagnetism
  in vanadium-oxide based perovskites},\ }\href
  {https://doi.org/10.1103/PhysRevB.87.155127} {\bibfield  {journal} {\bibinfo
  {journal} {Physical Review B}\ }\textbf {\bibinfo {volume} {87}},\ \bibinfo
  {pages} {155127} (\bibinfo {year} {2013}{\natexlab{b}})}\BibitemShut
  {NoStop}%
\bibitem [{\citenamefont {Kresse}\ and\ \citenamefont
  {Furthm{\"u}ller}(1996{\natexlab{a}})}]{kresse1996efficiency}%
  \BibitemOpen
  \bibfield  {author} {\bibinfo {author} {\bibfnamefont {G.}~\bibnamefont
  {Kresse}}\ and\ \bibinfo {author} {\bibfnamefont {J.}~\bibnamefont
  {Furthm{\"u}ller}},\ }\bibfield  {title} {\bibinfo {title} {Efficiency of
  ab-initio total energy calculations for metals and semiconductors using a
  plane-wave basis set},\ }\href
  {https://doi.org/https://doi.org/10.1016/0927-0256(96)00008-0} {\bibfield
  {journal} {\bibinfo  {journal} {Computational materials science}\ }\textbf
  {\bibinfo {volume} {6}},\ \bibinfo {pages} {15} (\bibinfo {year}
  {1996}{\natexlab{a}})}\BibitemShut {NoStop}%
\bibitem [{\citenamefont {Kresse}\ and\ \citenamefont
  {Furthm{\"u}ller}(1996{\natexlab{b}})}]{kresse1996efficient}%
  \BibitemOpen
  \bibfield  {author} {\bibinfo {author} {\bibfnamefont {G.}~\bibnamefont
  {Kresse}}\ and\ \bibinfo {author} {\bibfnamefont {J.}~\bibnamefont
  {Furthm{\"u}ller}},\ }\bibfield  {title} {\bibinfo {title} {Efficient
  iterative schemes for ab initio total-energy calculations using a plane-wave
  basis set},\ }\href {https://doi.org/10.1103/PhysRevB.54.11169} {\bibfield
  {journal} {\bibinfo  {journal} {Physical Review B}\ }\textbf {\bibinfo
  {volume} {54}},\ \bibinfo {pages} {11169} (\bibinfo {year}
  {1996}{\natexlab{b}})}\BibitemShut {NoStop}%
\bibitem [{\citenamefont {Bl{\"o}chl}(1994)}]{blochl1994projector}%
  \BibitemOpen
  \bibfield  {author} {\bibinfo {author} {\bibfnamefont {P.~E.}\ \bibnamefont
  {Bl{\"o}chl}},\ }\bibfield  {title} {\bibinfo {title} {Projector
  augmented-wave method},\ }\href {https://doi.org/10.1103/PhysRevB.50.17953}
  {\bibfield  {journal} {\bibinfo  {journal} {Physical Review B}\ }\textbf
  {\bibinfo {volume} {50}},\ \bibinfo {pages} {17953} (\bibinfo {year}
  {1994})}\BibitemShut {NoStop}%
\bibitem [{\citenamefont {Perdew}\ \emph {et~al.}(1996)\citenamefont {Perdew},
  \citenamefont {Burke},\ and\ \citenamefont
  {Ernzerhof}}]{perdew1996generalized}%
  \BibitemOpen
  \bibfield  {author} {\bibinfo {author} {\bibfnamefont {J.~P.}\ \bibnamefont
  {Perdew}}, \bibinfo {author} {\bibfnamefont {K.}~\bibnamefont {Burke}},\ and\
  \bibinfo {author} {\bibfnamefont {M.}~\bibnamefont {Ernzerhof}},\ }\bibfield
  {title} {\bibinfo {title} {Generalized gradient approximation made simple},\
  }\href {https://doi.org/10.1103/PhysRevLett.77.3865} {\bibfield  {journal}
  {\bibinfo  {journal} {Physical Review Letters}\ }\textbf {\bibinfo {volume}
  {77}},\ \bibinfo {pages} {3865} (\bibinfo {year} {1996})}\BibitemShut
  {NoStop}%
\bibitem [{\citenamefont {Dudarev}\ \emph {et~al.}(1998)\citenamefont
  {Dudarev}, \citenamefont {Botton}, \citenamefont {Savrasov}, \citenamefont
  {Humphreys},\ and\ \citenamefont {Sutton}}]{dudarev1998electron}%
  \BibitemOpen
  \bibfield  {author} {\bibinfo {author} {\bibfnamefont {S.~L.}\ \bibnamefont
  {Dudarev}}, \bibinfo {author} {\bibfnamefont {G.~A.}\ \bibnamefont {Botton}},
  \bibinfo {author} {\bibfnamefont {S.~Y.}\ \bibnamefont {Savrasov}}, \bibinfo
  {author} {\bibfnamefont {C.~J.}\ \bibnamefont {Humphreys}},\ and\ \bibinfo
  {author} {\bibfnamefont {A.~P.}\ \bibnamefont {Sutton}},\ }\bibfield  {title}
  {\bibinfo {title} {{Electron-energy-loss spectra and the structural stability
  of nickel oxide: An LSDA+U study}},\ }\href
  {https://doi.org/10.1103/PhysRevB.57.1505} {\bibfield  {journal} {\bibinfo
  {journal} {Physical Review B}\ }\textbf {\bibinfo {volume} {57}},\ \bibinfo
  {pages} {1505} (\bibinfo {year} {1998})}\BibitemShut {NoStop}%
\bibitem [{\citenamefont {Pentcheva}\ and\ \citenamefont
  {Pickett}(2008)}]{pentcheva2008ionic}%
  \BibitemOpen
  \bibfield  {author} {\bibinfo {author} {\bibfnamefont {R.}~\bibnamefont
  {Pentcheva}}\ and\ \bibinfo {author} {\bibfnamefont {W.~E.}\ \bibnamefont
  {Pickett}},\ }\bibfield  {title} {\bibinfo {title} {{Ionic relaxation
  contribution to the electronic reconstruction at the $n$-type
  ${\text{LaAlO}}_{3}/{\text{SrTiO}}_{3}$ interface}},\ }\href
  {https://doi.org/10.1103/PhysRevB.78.205106} {\bibfield  {journal} {\bibinfo
  {journal} {Physical Review B}\ }\textbf {\bibinfo {volume} {78}},\ \bibinfo
  {pages} {205106} (\bibinfo {year} {2008})}\BibitemShut {NoStop}%
\bibitem [{\citenamefont {Cooper}(2012)}]{cooper2012enhanced}%
  \BibitemOpen
  \bibfield  {author} {\bibinfo {author} {\bibfnamefont {V.~R.}\ \bibnamefont
  {Cooper}},\ }\bibfield  {title} {\bibinfo {title} {Enhanced carrier
  mobilities in two-dimensional electron gases at {III-III/I-V} oxide
  heterostructure interfaces},\ }\href
  {https://doi.org/10.1103/PhysRevB.85.235109} {\bibfield  {journal} {\bibinfo
  {journal} {Physical Review B}\ }\textbf {\bibinfo {volume} {85}},\ \bibinfo
  {pages} {235109} (\bibinfo {year} {2012})}\BibitemShut {NoStop}%
\bibitem [{\citenamefont {Assmann}\ \emph {et~al.}(2013)\citenamefont
  {Assmann}, \citenamefont {Blaha}, \citenamefont {Laskowski}, \citenamefont
  {Held}, \citenamefont {Okamoto},\ and\ \citenamefont
  {Sangiovanni}}]{assmann2013oxide}%
  \BibitemOpen
  \bibfield  {author} {\bibinfo {author} {\bibfnamefont {E.}~\bibnamefont
  {Assmann}}, \bibinfo {author} {\bibfnamefont {P.}~\bibnamefont {Blaha}},
  \bibinfo {author} {\bibfnamefont {R.}~\bibnamefont {Laskowski}}, \bibinfo
  {author} {\bibfnamefont {K.}~\bibnamefont {Held}}, \bibinfo {author}
  {\bibfnamefont {S.}~\bibnamefont {Okamoto}},\ and\ \bibinfo {author}
  {\bibfnamefont {G.}~\bibnamefont {Sangiovanni}},\ }\bibfield  {title}
  {\bibinfo {title} {Oxide heterostructures for efficient solar cells},\ }\href
  {https://doi.org/10.1103/PhysRevLett.110.078701} {\bibfield  {journal}
  {\bibinfo  {journal} {Physical Review Letters}\ }\textbf {\bibinfo {volume}
  {110}},\ \bibinfo {pages} {078701} (\bibinfo {year} {2013})}\BibitemShut
  {NoStop}%
\bibitem [{pat()}]{patel2023layer}%
  \BibitemOpen
  \bibfield  {title} {\bibinfo {title} {{Supplementary Materials for
  $\sqrt{2}\times\sqrt{2}$ supercell magnetic calculations. It also contains
  bands for KTO$_8$ layers, layer projected DOS for 4/4, atomic movements after
  relaxation, tight-binding bands, and also Tables for relative ground state
  energies and partial free charges.}},\ }\href@noop {} {\ }\BibitemShut
  {NoStop}%
\bibitem [{\citenamefont {Zhang}\ \emph {et~al.}(2023)\citenamefont {Zhang},
  \citenamefont {Pang}, \citenamefont {Yin}, \citenamefont {Yan}, \citenamefont
  {Lv}, \citenamefont {Deng},\ and\ \citenamefont {Zhang}}]{zhang2023quasi}%
  \BibitemOpen
  \bibfield  {author} {\bibinfo {author} {\bibfnamefont {X.}~\bibnamefont
  {Zhang}}, \bibinfo {author} {\bibfnamefont {Z.}~\bibnamefont {Pang}},
  \bibinfo {author} {\bibfnamefont {C.}~\bibnamefont {Yin}}, \bibinfo {author}
  {\bibfnamefont {M.}~\bibnamefont {Yan}}, \bibinfo {author} {\bibfnamefont
  {Y.-Y.}\ \bibnamefont {Lv}}, \bibinfo {author} {\bibfnamefont
  {Y.}~\bibnamefont {Deng}},\ and\ \bibinfo {author} {\bibfnamefont {S.-T.}\
  \bibnamefont {Zhang}},\ }\bibfield  {title} {\bibinfo {title}
  {Quasi-two-dimensional electron gas and weak antilocalization at the
  interface of {SrTaO$_3$/KTaO$_3$} heterostructures},\ }\href
  {https://doi.org/10.1103/PhysRevB.108.235114} {\bibfield  {journal} {\bibinfo
   {journal} {Physical Review B}\ }\textbf {\bibinfo {volume} {108}},\ \bibinfo
  {pages} {235114} (\bibinfo {year} {2023})}\BibitemShut {NoStop}%
\bibitem [{\citenamefont {Sawada}\ \emph {et~al.}(1996)\citenamefont {Sawada},
  \citenamefont {Hamada}, \citenamefont {Terakura},\ and\ \citenamefont
  {Asada}}]{sawada1996orbital}%
  \BibitemOpen
  \bibfield  {author} {\bibinfo {author} {\bibfnamefont {H.}~\bibnamefont
  {Sawada}}, \bibinfo {author} {\bibfnamefont {N.}~\bibnamefont {Hamada}},
  \bibinfo {author} {\bibfnamefont {K.}~\bibnamefont {Terakura}},\ and\
  \bibinfo {author} {\bibfnamefont {T.}~\bibnamefont {Asada}},\ }\bibfield
  {title} {\bibinfo {title} {Orbital and spin orderings in
  {${\mathrm{YVO}}_{3}$ and ${\mathrm{LaVO}}_{3}$} in the generalized gradient
  approximation},\ }\href {https://doi.org/10.1103/PhysRevB.53.12742}
  {\bibfield  {journal} {\bibinfo  {journal} {Physical Review B}\ }\textbf
  {\bibinfo {volume} {53}},\ \bibinfo {pages} {12742} (\bibinfo {year}
  {1996})}\BibitemShut {NoStop}%
\bibitem [{\citenamefont {Fang}\ \emph {et~al.}(2003)\citenamefont {Fang},
  \citenamefont {Nagaosa},\ and\ \citenamefont
  {Terakura}}]{fang2003anisotropic}%
  \BibitemOpen
  \bibfield  {author} {\bibinfo {author} {\bibfnamefont {Z.}~\bibnamefont
  {Fang}}, \bibinfo {author} {\bibfnamefont {N.}~\bibnamefont {Nagaosa}},\ and\
  \bibinfo {author} {\bibfnamefont {K.}~\bibnamefont {Terakura}},\ }\bibfield
  {title} {\bibinfo {title} {{Anisotropic optical conductivities due to spin
  and orbital ordering in ${\mathrm{LaVO}}_{3}$ and ${\mathrm{YVO}}_{3}:$
  First-principles studies}},\ }\href
  {https://doi.org/10.1103/PhysRevB.67.035101} {\bibfield  {journal} {\bibinfo
  {journal} {Physical Review B}\ }\textbf {\bibinfo {volume} {67}},\ \bibinfo
  {pages} {035101} (\bibinfo {year} {2003})}\BibitemShut {NoStop}%
\bibitem [{\citenamefont {Fang}\ and\ \citenamefont
  {Nagaosa}(2004)}]{fang2004quantum}%
  \BibitemOpen
  \bibfield  {author} {\bibinfo {author} {\bibfnamefont {Z.}~\bibnamefont
  {Fang}}\ and\ \bibinfo {author} {\bibfnamefont {N.}~\bibnamefont {Nagaosa}},\
  }\bibfield  {title} {\bibinfo {title} {{Quantum Versus Jahn-Teller Orbital
  Physics in ${\mathrm{Y}\mathrm{V}\mathrm{O}}_{3}$ and
  ${\mathrm{L}\mathrm{a}\mathrm{V}\mathrm{O}}_{3}$}},\ }\href
  {https://doi.org/10.1103/PhysRevLett.93.176404} {\bibfield  {journal}
  {\bibinfo  {journal} {Physical Review Letters}\ }\textbf {\bibinfo {volume}
  {93}},\ \bibinfo {pages} {176404} (\bibinfo {year} {2004})}\BibitemShut
  {NoStop}%
\bibitem [{\citenamefont {Mahajan}\ \emph {et~al.}(1992)\citenamefont
  {Mahajan}, \citenamefont {Johnston}, \citenamefont {Torgeson},\ and\
  \citenamefont {Borsa}}]{mahajan1992magnetic}%
  \BibitemOpen
  \bibfield  {author} {\bibinfo {author} {\bibfnamefont {A.~V.}\ \bibnamefont
  {Mahajan}}, \bibinfo {author} {\bibfnamefont {D.~C.}\ \bibnamefont
  {Johnston}}, \bibinfo {author} {\bibfnamefont {D.~R.}\ \bibnamefont
  {Torgeson}},\ and\ \bibinfo {author} {\bibfnamefont {F.}~\bibnamefont
  {Borsa}},\ }\bibfield  {title} {\bibinfo {title} {{Magnetic properties of
  ${\mathrm{LaVO}}_{3}$}},\ }\href {https://doi.org/10.1103/PhysRevB.46.10966}
  {\bibfield  {journal} {\bibinfo  {journal} {Physical Review B}\ }\textbf
  {\bibinfo {volume} {46}},\ \bibinfo {pages} {10966} (\bibinfo {year}
  {1992})}\BibitemShut {NoStop}%
\bibitem [{\citenamefont {de~Groot}\ \emph {et~al.}(1983)\citenamefont
  {de~Groot}, \citenamefont {Mueller}, \citenamefont {Engen},\ and\
  \citenamefont {Buschow}}]{de1983new}%
  \BibitemOpen
  \bibfield  {author} {\bibinfo {author} {\bibfnamefont {R.~A.}\ \bibnamefont
  {de~Groot}}, \bibinfo {author} {\bibfnamefont {F.~M.}\ \bibnamefont
  {Mueller}}, \bibinfo {author} {\bibfnamefont {P.~G.~v.}\ \bibnamefont
  {Engen}},\ and\ \bibinfo {author} {\bibfnamefont {K.~H.~J.}\ \bibnamefont
  {Buschow}},\ }\bibfield  {title} {\bibinfo {title} {{New Class of Materials:
  Half-Metallic Ferromagnets}},\ }\href
  {https://doi.org/10.1103/PhysRevLett.50.2024} {\bibfield  {journal} {\bibinfo
   {journal} {Physical Review Letters}\ }\textbf {\bibinfo {volume} {50}},\
  \bibinfo {pages} {2024} (\bibinfo {year} {1983})}\BibitemShut {NoStop}%
\bibitem [{\citenamefont {Picozzi}\ \emph {et~al.}(2003)\citenamefont
  {Picozzi}, \citenamefont {Continenza},\ and\ \citenamefont
  {Freeman}}]{picozzi2003first}%
  \BibitemOpen
  \bibfield  {author} {\bibinfo {author} {\bibfnamefont {S.}~\bibnamefont
  {Picozzi}}, \bibinfo {author} {\bibfnamefont {A.}~\bibnamefont
  {Continenza}},\ and\ \bibinfo {author} {\bibfnamefont {A.}~\bibnamefont
  {Freeman}},\ }\bibfield  {title} {\bibinfo {title} {{First principles study
  of electronic and magnetic properties of Co2MnGe/GaAs interfaces}},\ }\href
  {https://doi.org/10.1016/S0022-3697(03)00121-5} {\bibfield  {journal}
  {\bibinfo  {journal} {Journal of Physics and Chemistry of Solids}\ }\textbf
  {\bibinfo {volume} {64}},\ \bibinfo {pages} {1697} (\bibinfo {year}
  {2003})}\BibitemShut {NoStop}%
\bibitem [{\citenamefont {Mavropoulos}\ \emph {et~al.}(2005)\citenamefont
  {Mavropoulos}, \citenamefont {Le{\v{z}}ai{\'c}},\ and\ \citenamefont
  {Bl\"ugel}}]{mavropoulos2005half}%
  \BibitemOpen
  \bibfield  {author} {\bibinfo {author} {\bibfnamefont {P.}~\bibnamefont
  {Mavropoulos}}, \bibinfo {author} {\bibfnamefont {M.}~\bibnamefont
  {Le{\v{z}}ai{\'c}}},\ and\ \bibinfo {author} {\bibfnamefont {S.}~\bibnamefont
  {Bl\"ugel}},\ }\bibfield  {title} {\bibinfo {title} {Half-metallic
  ferromagnets for magnetic tunnel junctions by ab initio calculations},\
  }\href {https://doi.org/10.1103/PhysRevB.72.174428} {\bibfield  {journal}
  {\bibinfo  {journal} {Physical Review B}\ }\textbf {\bibinfo {volume} {72}},\
  \bibinfo {pages} {174428} (\bibinfo {year} {2005})}\BibitemShut {NoStop}%
\bibitem [{\citenamefont {Hashemifar}\ \emph {et~al.}(2005)\citenamefont
  {Hashemifar}, \citenamefont {Kratzer},\ and\ \citenamefont
  {Scheffler}}]{hashemifar2005preserving}%
  \BibitemOpen
  \bibfield  {author} {\bibinfo {author} {\bibfnamefont {S.~J.}\ \bibnamefont
  {Hashemifar}}, \bibinfo {author} {\bibfnamefont {P.}~\bibnamefont
  {Kratzer}},\ and\ \bibinfo {author} {\bibfnamefont {M.}~\bibnamefont
  {Scheffler}},\ }\bibfield  {title} {\bibinfo {title} {{Preserving the
  Half-Metallicity at the Heusler Alloy
  ${\mathrm{C}\mathrm{o}}_{2}\mathrm{M}\mathrm{n}\mathrm{S}\mathrm{i}(001)$
  Surface: A Density Functional Theory Study}},\ }\href
  {https://doi.org/10.1103/PhysRevLett.94.096402} {\bibfield  {journal}
  {\bibinfo  {journal} {Physical Review Letters}\ }\textbf {\bibinfo {volume}
  {94}},\ \bibinfo {pages} {096402} (\bibinfo {year} {2005})}\BibitemShut
  {NoStop}%
\bibitem [{\citenamefont {Di~Marco}\ \emph {et~al.}(2018)\citenamefont
  {Di~Marco}, \citenamefont {Held}, \citenamefont {Keshavarz}, \citenamefont
  {Kvashnin},\ and\ \citenamefont {Chioncel}}]{di2018half}%
  \BibitemOpen
  \bibfield  {author} {\bibinfo {author} {\bibfnamefont {I.}~\bibnamefont
  {Di~Marco}}, \bibinfo {author} {\bibfnamefont {A.}~\bibnamefont {Held}},
  \bibinfo {author} {\bibfnamefont {S.}~\bibnamefont {Keshavarz}}, \bibinfo
  {author} {\bibfnamefont {Y.~O.}\ \bibnamefont {Kvashnin}},\ and\ \bibinfo
  {author} {\bibfnamefont {L.}~\bibnamefont {Chioncel}},\ }\bibfield  {title}
  {\bibinfo {title} {{Half-metallicity and magnetism in the
  ${\mathrm{Co}}_{2}\mathrm{MnAl}/\mathrm{CoMnVAl}$ heterostructure}},\ }\href
  {https://doi.org/10.1103/PhysRevB.97.035105} {\bibfield  {journal} {\bibinfo
  {journal} {Physical Review B}\ }\textbf {\bibinfo {volume} {97}},\ \bibinfo
  {pages} {035105} (\bibinfo {year} {2018})}\BibitemShut {NoStop}%
\bibitem [{\citenamefont {Jena}\ and\ \citenamefont
  {Datta}(2023)}]{jena2023evidence}%
  \BibitemOpen
  \bibfield  {author} {\bibinfo {author} {\bibfnamefont {S.}~\bibnamefont
  {Jena}}\ and\ \bibinfo {author} {\bibfnamefont {S.}~\bibnamefont {Datta}},\
  }\bibfield  {title} {\bibinfo {title} {Evidence of half-metallicity at the
  {BiFeO$_3$} (001) surface},\ }\href {https://doi.org/10.1039/D2NJ06169D}
  {\bibfield  {journal} {\bibinfo  {journal} {New Journal of Chemistry}\
  }\textbf {\bibinfo {volume} {47}},\ \bibinfo {pages} {6983} (\bibinfo {year}
  {2023})}\BibitemShut {NoStop}%
\bibitem [{\citenamefont {Michael~Ziese}(2001)}]{spinelectronics}%
  \BibitemOpen
  \bibfield  {author} {\bibinfo {author} {\bibfnamefont {M.~J.~T.}\
  \bibnamefont {Michael~Ziese}},\ }\href
  {https://doi.org/10.1007/3-540-45258-3} {\emph {\bibinfo {title} {Spin
  Electronics}}},\ Lecture Notes in Physics\ (\bibinfo  {publisher} {Springer
  Berlin, Heidelberg},\ \bibinfo {year} {2001})\BibitemShut {NoStop}%
\bibitem [{\citenamefont {Zhang}\ \emph {et~al.}(2020)\citenamefont {Zhang},
  \citenamefont {Zhao}, \citenamefont {Gautreau}, \citenamefont {Raczkowski},
  \citenamefont {Saha}, \citenamefont {Garlea}, \citenamefont {Cao},
  \citenamefont {Hong}, \citenamefont {Jeschke}, \citenamefont {Mahanti},
  \citenamefont {Birol}, \citenamefont {Assaad},\ and\ \citenamefont
  {Ke}}]{zhang2020coexistence}%
  \BibitemOpen
  \bibfield  {author} {\bibinfo {author} {\bibfnamefont {H.}~\bibnamefont
  {Zhang}}, \bibinfo {author} {\bibfnamefont {Z.}~\bibnamefont {Zhao}},
  \bibinfo {author} {\bibfnamefont {D.}~\bibnamefont {Gautreau}}, \bibinfo
  {author} {\bibfnamefont {M.}~\bibnamefont {Raczkowski}}, \bibinfo {author}
  {\bibfnamefont {A.}~\bibnamefont {Saha}}, \bibinfo {author} {\bibfnamefont
  {V.~O.}\ \bibnamefont {Garlea}}, \bibinfo {author} {\bibfnamefont
  {H.}~\bibnamefont {Cao}}, \bibinfo {author} {\bibfnamefont {T.}~\bibnamefont
  {Hong}}, \bibinfo {author} {\bibfnamefont {H.~O.}\ \bibnamefont {Jeschke}},
  \bibinfo {author} {\bibfnamefont {S.~D.}\ \bibnamefont {Mahanti}}, \bibinfo
  {author} {\bibfnamefont {T.}~\bibnamefont {Birol}}, \bibinfo {author}
  {\bibfnamefont {F.~F.}\ \bibnamefont {Assaad}},\ and\ \bibinfo {author}
  {\bibfnamefont {X.}~\bibnamefont {Ke}},\ }\bibfield  {title} {\bibinfo
  {title} {Coexistence and interaction of spinons and magnons in an
  antiferromagnet with alternating antiferromagnetic and ferromagnetic quantum
  spin chains},\ }\href {https://doi.org/10.1103/PhysRevLett.125.037204}
  {\bibfield  {journal} {\bibinfo  {journal} {Physical Review Letters}\
  }\textbf {\bibinfo {volume} {125}},\ \bibinfo {pages} {037204} (\bibinfo
  {year} {2020})}\BibitemShut {NoStop}%
\bibitem [{\citenamefont {Anderson}(1950)}]{anderson1950antiferromagnetism}%
  \BibitemOpen
  \bibfield  {author} {\bibinfo {author} {\bibfnamefont {P.~W.}\ \bibnamefont
  {Anderson}},\ }\bibfield  {title} {\bibinfo {title} {Antiferromagnetism.
  theory of superexchange interaction},\ }\href
  {https://doi.org/10.1103/PhysRev.79.350} {\bibfield  {journal} {\bibinfo
  {journal} {Physical Review}\ }\textbf {\bibinfo {volume} {79}},\ \bibinfo
  {pages} {350} (\bibinfo {year} {1950})}\BibitemShut {NoStop}%
\bibitem [{\citenamefont {Goodenough}\ and\ \citenamefont
  {Loeb}(1955)}]{goodenough1954theory}%
  \BibitemOpen
  \bibfield  {author} {\bibinfo {author} {\bibfnamefont {J.~B.}\ \bibnamefont
  {Goodenough}}\ and\ \bibinfo {author} {\bibfnamefont {A.~L.}\ \bibnamefont
  {Loeb}},\ }\bibfield  {title} {\bibinfo {title} {Theory of ionic ordering,
  crystal distortion, and magnetic exchange due to covalent forces in
  spinels},\ }\href {https://doi.org/10.1103/PhysRev.98.391} {\bibfield
  {journal} {\bibinfo  {journal} {Physical Review}\ }\textbf {\bibinfo {volume}
  {98}},\ \bibinfo {pages} {391} (\bibinfo {year} {1955})}\BibitemShut
  {NoStop}%
\bibitem [{\citenamefont {Goodenough}(1955)}]{goodenough1955theory}%
  \BibitemOpen
  \bibfield  {author} {\bibinfo {author} {\bibfnamefont {J.~B.}\ \bibnamefont
  {Goodenough}},\ }\bibfield  {title} {\bibinfo {title} {Theory of the role of
  covalence in the perovskite-type manganites {[La,M(II)]MnO$_3$}},\ }\href
  {https://doi.org/10.1103/PhysRev.100.564} {\bibfield  {journal} {\bibinfo
  {journal} {Physical Review}\ }\textbf {\bibinfo {volume} {100}},\ \bibinfo
  {pages} {564} (\bibinfo {year} {1955})}\BibitemShut {NoStop}%
\bibitem [{\citenamefont {Kanamori}(1959)}]{kanamori1959superexchange}%
  \BibitemOpen
  \bibfield  {author} {\bibinfo {author} {\bibfnamefont {J.}~\bibnamefont
  {Kanamori}},\ }\bibfield  {title} {\bibinfo {title} {Superexchange
  interaction and symmetry properties of electron orbitals},\ }\href
  {https://doi.org/https://doi.org/10.1016/0022-3697(59)90061-7} {\bibfield
  {journal} {\bibinfo  {journal} {Journal of Physics and Chemistry of Solids}\
  }\textbf {\bibinfo {volume} {10}},\ \bibinfo {pages} {87} (\bibinfo {year}
  {1959})}\BibitemShut {NoStop}%
\bibitem [{\citenamefont {Wang}\ and\ \citenamefont
  {Chang}(2022)}]{wang2022goodenough}%
  \BibitemOpen
  \bibfield  {author} {\bibinfo {author} {\bibfnamefont {M.-C.}\ \bibnamefont
  {Wang}}\ and\ \bibinfo {author} {\bibfnamefont {C.-R.}\ \bibnamefont
  {Chang}},\ }\bibfield  {title} {\bibinfo {title}
  {{Goodenough-Kanamori-Anderson Rules in {CrI$_3$/MoTe$_2$/CrI$_3$} Van der
  Waals Heterostructure}},\ }\href {https://doi.org/10.1149/1945-7111/ac7006}
  {\bibfield  {journal} {\bibinfo  {journal} {Journal of The Electrochemical
  Society}\ }\textbf {\bibinfo {volume} {169}},\ \bibinfo {pages} {053507}
  (\bibinfo {year} {2022})}\BibitemShut {NoStop}%
\bibitem [{\citenamefont {King-Smith}\ and\ \citenamefont
  {Vanderbilt}(1993)}]{king1993theory}%
  \BibitemOpen
  \bibfield  {author} {\bibinfo {author} {\bibfnamefont {R.~D.}\ \bibnamefont
  {King-Smith}}\ and\ \bibinfo {author} {\bibfnamefont {D.}~\bibnamefont
  {Vanderbilt}},\ }\bibfield  {title} {\bibinfo {title} {Theory of polarization
  of crystalline solids},\ }\href {https://doi.org/10.1103/PhysRevB.47.1651}
  {\bibfield  {journal} {\bibinfo  {journal} {Physical Review B}\ }\textbf
  {\bibinfo {volume} {47}},\ \bibinfo {pages} {1651} (\bibinfo {year}
  {1993})}\BibitemShut {NoStop}%
\bibitem [{\citenamefont {Stengel}\ and\ \citenamefont
  {Vanderbilt}(2009)}]{stengel2009berry}%
  \BibitemOpen
  \bibfield  {author} {\bibinfo {author} {\bibfnamefont {M.}~\bibnamefont
  {Stengel}}\ and\ \bibinfo {author} {\bibfnamefont {D.}~\bibnamefont
  {Vanderbilt}},\ }\bibfield  {title} {\bibinfo {title} {Berry-phase theory of
  polar discontinuities at oxide-oxide interfaces},\ }\href
  {https://doi.org/10.1103/PhysRevB.80.241103} {\bibfield  {journal} {\bibinfo
  {journal} {Physical Review B}\ }\textbf {\bibinfo {volume} {80}},\ \bibinfo
  {pages} {241103} (\bibinfo {year} {2009})}\BibitemShut {NoStop}%
\bibitem [{\citenamefont {Stengel}(2011)}]{stengel2011first}%
  \BibitemOpen
  \bibfield  {author} {\bibinfo {author} {\bibfnamefont {M.}~\bibnamefont
  {Stengel}},\ }\bibfield  {title} {\bibinfo {title} {First-principles modeling
  of electrostatically doped perovskite systems},\ }\href
  {https://doi.org/10.1103/PhysRevLett.106.136803} {\bibfield  {journal}
  {\bibinfo  {journal} {Physical Review Letters}\ }\textbf {\bibinfo {volume}
  {106}},\ \bibinfo {pages} {136803} (\bibinfo {year} {2011})}\BibitemShut
  {NoStop}%
\end{thebibliography}%
\end{document}